\documentclass[onecolumn,amsmath,amssymb,nofootinbib,12pt]{article}
\pdfoutput=1
\usepackage{jheppub}
\usepackage{graphicx}% Include figure files
\usepackage{dcolumn}% Align table columns on decimal point
\usepackage{bm,psfrag}% bold math

\usepackage{subfigure}
\usepackage{float}
\usepackage{tensor}

\newcommand{\ben}{\begin{equation}}
\newcommand{\een}{\end{equation}}
\newcommand{\be}{\begin{equation}}
\newcommand{\ee}{\end{equation}}
\newcommand{\bea}{\begin{eqnarray}}
\newcommand{\eea}{\end{eqnarray}}
\newcommand{\ba}{\begin{eqnarray}}
\newcommand{\ea}{\end{eqnarray}}

\newcommand{\beq}{\begin{equation}}
\newcommand{\eeq}{\end{equation}}
\newcommand{\beqa}{\begin{eqnarray}}
\newcommand{\eeqa}{\end{eqnarray}}
\newcommand{\beqar}{\begin{eqnarray*}}
\newcommand{\eeqar}{\end{eqnarray*}}

\newcommand{\reef}[1]{(\ref{#1})}

\newcommand{\eg}{{\it e.g.,}\ }
\newcommand{\ie}{{\it i.e.,}\ }

\newcommand{\labell}[1]{\label{#1}} %{\mt{#1}\label{#1}} %

\def\t6 {T_\mt{D6}}

\newcommand{\mt}[1]{\textrm{\tiny #1}}

\newcommand{\vk}{{\vec{k}}}
\newcommand{\vx}{{\vec{x}}}

\def\cale         {{\cal E}}
\def\ce {{\cal E}}

\def\calo         {{\cal O}}

\def\ee           {{\rm e}}

\def\sqr#1#2{{\vcenter{\vbox{\hrule height.#2pt
 \hbox{\vrule width.#2pt height#1pt \kern#1pt
 \vrule width.#2pt}\hrule height.#2pt}}}}

\def\w{\omega}

\def\ee{\cale}

\def\aa1{\phi}
\def\cc1{\psi}

\def\vev#1{\langle #1 \rangle}

\def\lzero{\lambda_0}

\def\nnn{\nonumber}

\newcommand{\dt}{\delta t}

\begin{document}

\preprint{arXiv:1505.05224 [hep-th]}

\title{Smooth and fast versus instantaneous quenches in quantum
  field theory}

\author{Sumit R. Das,$^{1}$ Dami\'an A. Galante$^{2,3,4}$ and Robert C. Myers$^3$}
\affiliation{$^1$\,Department of Physics and Astronomy, University of Kentucky,\\ 
\vphantom{k}\ \ Lexington, KY 40506, USA}
\affiliation{$^2$\,Department of Applied Mathematics, University of Western Ontario,\\ 
\vphantom{k}\ \ London, ON N6A 5B7, Canada}
\affiliation{$^3$\,Perimeter Institute for Theoretical Physics, Waterloo, ON N2L 2Y5, Canada}
\affiliation{$^4$\,Kavli Institute for Theoretical Physics, University of California, \\
\vphantom{k}\ \ Santa Barbara, CA 93106, USA}

\emailAdd{das@pa.uky.edu}
\emailAdd{dgalante@perimeterinstitute.ca}
\emailAdd{rmyers@perimeterinstitute.ca}

\date{\today}

\abstract{We examine in detail the relationship between smooth fast
  quantum quenches, characterized by a time scale $\dt$, and {\em
    instantaneous quenches}, within the framework of exactly solvable mass quenches in free
  scalar field theory. Our earlier studies \cite{dgm1,dgm2} highlighted that the two protocols remain distinct in the limit $\dt\to0$ because of the relation of the quench rate to the UV cut-off, \ie $1/\dt\ll\Lambda$ always holds in the fast smooth quenches while $1/\dt\sim\Lambda$ for instantaneous quenches. Here we study UV finite quantities like correlators at finite spatial distances and the excess energy produced above the final ground state energy. We show that at late times and large distances (compared to the quench time scale) the smooth quench correlator approaches that for the instantaneous quench. At early times, we find that for small spatial separation and small $\dt$, the correlator scales universally with $\dt$, exactly as in the scaling of renormalized one point functions found in earlier work. At larger separation, the dependence on $\dt$ drops out. The excess energy density is finite (for finite $m\dt$) and scales in a universal fashion for all $d$. However, the scaling behaviour produces a divergent result in the limit $m\dt \rightarrow 0$ for $d\ge4$, just as in an instantaneous quench, where it is UV divergent for $d \geq 4$.  We argue that similar results hold for arbitrary interacting theories: the excess energy density produced is expected to diverge for scaling dimensions $\Delta > d/2$.}

%\pacs{Valid PACS appear here}% PACS, the Physics and Astronomy
                             % Classification Scheme.
%\keywords{Suggested keywords}%Use showkeys class option if keyword
                              %display desired
\maketitle

%\tableofcontents

\newpage

\section{Introduction}
Universal scaling behavior in systems undergoing a quantum (or thermal) quench which involves critical points have been a subject of great interest in recent years \cite{more,kibble,zurek,qcritkz}. The classic example of such behavior is Kibble-Zurek scaling \cite{kibble,zurek} which involves a quench which starts from a gapped phase at a rate which is {\em slow} compared to the scale set by the initial gap. At the other extreme, there are a different set of universal behaviors in two-dimensional field theories which are quenched {\em instantaneously} from a gapped phase to a critical point \cite{cc2,cc3} and for instantaneous quenches which can be treated perturbatively \cite{gritsev}.

The AdS/CFT correspondence has yielded important insight in this area, both for Kibble-Zurek scaling \cite{holo-kz} and for novel non-equilibrium phases \cite{holo-sc}. Perhaps more significantly, holographic studies have led to the discovery of {\em new} scaling behavior for smooth quenches which are {\em fast compared to the physical mass scales, but slow compared to the UV scale} \cite{numer,fastQ}. In \cite{dgm1} and \cite{dgm2} we argued that this scaling law holds regardless of holography, and is valid for time dependent relevant deformations of generic conformal field theories (see also \cite{david}). Consider an action
\ben
S = S_{CFT} + \int d^d x~\lambda (t) \, \calo_\Delta (\vx,t)
\label{0-1}
\een
where the conformal dimension of the operator $\calo$ is $\Delta$ and $\lambda(t)$ interpolates between the constant values $\lambda_1$ and $\lambda_2$ (with an amplitude variation of $\delta\lambda$) over a time of order $\dt$. Then in the fast quench limit
\ben
\delta t \ll \lambda_1^{1/(\Delta-d)}, \lambda_2^{1/(\Delta-d)}, \delta \lambda^{1/(\Delta-d)}
\label{0-2}
\een
the renormalized energy density $\delta \cale_{ren}$ scales as 
\ben
\delta \cale_{ren} \sim  \frac{\delta \lambda^2}{\delta t^{2\Delta-d}}\,.
\label{0-3}
\een
Similarly, the peak of the renormalized expectation value of the quenched operator,
measured at times earlier than or soon after the end of the quench, was also found to scale as 
\ben
\langle \calo_\Delta \rangle_{ren} \sim \frac{\delta \lambda}{\delta t^{2\Delta-d}} \, ,
\label{0-4}
\een
This general result emerged out of detailed investigations of exactly solvable mass quenches in free bosonic and fermionic theories. One important outcome of our analysis was an understanding of the relationship between smooth fast quenches for small $\dt$ and the instantaneous quenches of \eg \cite{cc2,cc3}. The latter involve a quench rate which is fast compared to all scales, including the UV cutoff, while smooth quench rates are faster than any physical scales, but slower than the cutoff scale.  On the other hand, {\em local} quantities like the energy density or $\langle\calo\rangle$ involve a sum over all momenta all the way to the cutoff --- for such quantities one does not expect the smooth quench result to agree with those in the instantaneous quench. By the same token, one would expect that for correlators at finite separations larger than $\dt$, there should be agreement. In \cite{dgm1,dgm2} we also explored if the {\em late time} behavior of local quantities also agree, finding agreement at least in the $d=3$ case.

In this paper we continue to explore the relationship between fast but smooth quenches and instantaneous  quenches in further detail. Our analysis will focus on quenches in free scalar field theory with a time-dependent mass. However we argue that the lessons we draw there will be valid for quenches in interacting theories of the type described above. 

The scaling laws in \cite{dgm1,dgm2} were derived for {\em
  renormalized} composite operators, which are the appropriate
quantities for quench rates much slower than the cutoff scale. In this
work, we examine the late time behavior of such operators. In
addition, we
focus on quantities which are UV finite, \eg two-point correlation functions at finite spatial separations and the excess energy over the ground state energy at late times. 

In section \ref{late}, we consider late time correlators, $t \gg \dt$. We will show that for (suitably defined) large spatial separations, these correlators agree with the correlators for an instantaneous quench. For separations $r$ which are very small, \ie $mr \ll 1$ there is once again agreement, reflecting the fact that the dominant singular behavior for small separation is independent of any time dependence of the mass. The corrections to this leading small distance behavior are in one-to-one correspondence with the counterterms necessary to renormalize the composite operator $\phi^2$. In particular, the subleading small distance divergences involve derivatives of the mass function for $d \geq 6$. For intermediate separations, the two quench protocols lead to genuinely different results. 

In section \ref{uni_scaling}, we turn our attention to correlators at finite times $t \sim \dt$ and show that for $r\dt \gg 1$ the correlator becomes independent of $\dt$ as expected. For $ m^{-1} > \dt > r$ we find that the correlator exhibits a scaling behavior identical to that of the renormalized local operator $\langle\phi^2\rangle$. 

In section \ref{late2}, we consider the renormalized local quantity $\langle\phi^2\rangle$ at {\em late} times. We show that this quantity agrees for both quench protocols only for $d=3$. For $d=5$ and finite $\dt$, there is a slight difference between the smooth and the instantaneous answer in the limit of $\dt\to0$, while for the instantaneous quench, $\vev{\phi^2}$ is UV divergent for $d>5$.

In section \ref{energylate}, we consider the difference of the energy at late times and the ground state energy with the final value of the mass. This is one measure of the excess energy produced during the quench. We show that this quantity is explicitly UV finite. For $d \leq 3$ this becomes independent of $\dt$, in the $m\dt \ll 1$ limit. The next order correction, which scales as a power of $\dt$, is identical to the behavior of the {\em renormalized} energy in \cite{dgm1,dgm2}. For $d \geq 4$, the energy diverges in the $\dt \rightarrow  0$ limit, in the same way as the renormalized energy considered in \cite{dgm1,dgm2}.

In section \ref{general}, we discuss the validity of our results for the excess energy produced for general interacting field theories.
 
In section \ref{discuss}, we conclude with a brief discussion of our results and also consider various possible measures of the energy produced by the quench and their relationship.

\section{Bogoliubov coefficients for smooth and instantaneous quenches} \label{responseq}
Consider a scalar field in $d$ space-time dimensions with a time dependent mass,
\ben
S = - \int dt \int d^{d-1}x~\frac{1}{2}\,\left[ (\partial \phi)^2 + m^2(t) \phi^2 \right] \,.
\een
This theory is exactly solvable for a variety of different mass profiles, as described in \cite{dgm1,dgm2}. The quench protocol which we focus on here, involves the mass going from an initial value $m$ to zero at late times over a time scale $\dt$ with the smooth profile
\beq
m^2(t) = m^2\, \left( \frac{1 - \tanh (t/\dt)}{2} \right) \,.
\labell{massprofile}
\eeq
To solve the Klein-Gordon equation, we decompose the scalar field into momentum modes
\beq
\phi= \int\!\! \frac{d^{d-1}k}{(2\pi)^{(d-1)/2}} \ \left( a_\vk\, u_\vk + a^\dagger_\vk\, u^*_\vk\right)\,,\qquad
{\rm where}\ \ \ [a_\vk , a^\dagger_{\vk^\prime} ] = \delta^{d-1}(\vk - \vk^\prime)\,.
\labell{fieldx}
\eeq
The exact solution of the field equation is given by 
\cite{dgm1,dgm2,BD2}
\begin{eqnarray}
u_\vk & = & \frac{1}{\sqrt{2 \omega_{in}}} \exp(i\vk\cdot\vec{x}-i\omega_+ t - i\omega_- \dt \log (2 \cosh t/\dt)) \times \label{inmodes} \\
& &\qquad_2F_1 \left( 1+ i \omega_- \dt, i \omega_- \dt; 1 - i \omega_{in} \dt; \frac{1+\tanh(t/\dt)}{2} \right)\,,
\end{eqnarray}
where $_2F_1$ is the usual hypergeometric function and
\begin{eqnarray}
\omega_{in}  =  \sqrt{\vk^2+m^2} \,\,\, &,& \,\,\,  \omega_{out}   = |\vk|\,, \label{omegadef} \\
\omega_{\pm}   =  (\omega_{out} & \pm & \omega_{in})/2\,. \nonumber
\nonumber
\end{eqnarray}
The modes $u_\vk$ are the ``in" modes: they behave as plane waves in the infinite past and the $a_\vk$ annihilate the in-vacuum, \ie $a_\vk |in,0\rangle=0$. There is also another set of modes, the ``out-modes", which become plane waves in the infinite future,
\begin{eqnarray}
v_\vk  & = & \frac{1}{\sqrt{2\omega_{out}}} \exp(i\vk\cdot\vec{x}-i\omega_+ t - i\omega_- \dt \log (2 \cosh t/\dt)) \times \nnn \\
& & _2F_1 \left( 1+ i \omega_- \dt, i \omega_- \dt; 1 + i \omega_{out} \dt; \frac{1-\tanh(t/\dt)}{2} \right) \,.
\label{outmodes}
\end{eqnarray}
In terms of these, the field operator is
\beq
\phi= \int\!\! \frac{d^{d-1}k}{(2\pi)^{(d-1)/2}} \ \left( b_\vk\, v_\vk + b^\dagger_\vk\, v^*_\vk\right)\,,\qquad
{\rm where}\ \ \ [b_\vk , b^\dagger_{\vk^\prime} ] = \delta^{d-1}(\vk - \vk^\prime)\,.
\labell{fieldx2}
\eeq
%As in a vast part of the paper we will be interesting in the late time behaviour after the quench, it will be sometimes useful to use this basis. Of course they are both related by a Bogoliubov transformation. If we call $b_\vk$ the set of modes annihilating the out-vacuum, then the scalar field can be re-written as
%\beq
%\phi= \int\!\! \frac{d^{d-1}k}{(2\pi)^{d-1}}\ \left( b_\vk\, v_\vk + b^\dagger_\vk\, v^*_\vk\right)\,,
%\labell{fieldx2}
%\eeq
%where
%with the various frequencies defined in eq.~\reef{omegadef}. Note that now the $v_\vk$ behave as plane waves in the limit of $t\gg\dt$, \ie in the infinite future and {\it{not}} in the infinite past.
The Bogoliubov transformation that relates these two sets of modes is given by \cite{BD2}
\bea
u_\vk & = &\alpha_\vk\  v_\vk + \beta_\vk\  v^\star_{-\vk}, \nnn \\
u^\star_\vk & = &\alpha^\star_\vk\ v^\star_\vk + \beta^\star_\vk\ v_{-\vk},
\label{bogo}
\eea
where
\bea
\alpha_\vk & = & \sqrt{\frac{\omega_{out}}{\omega_{in}}} \, \frac{\Gamma (1-i\omega_{in}\delta t)\Gamma(-i\omega_{out}\delta t)}{\Gamma(-i\omega_+\delta t)\Gamma(1-i\omega_+\delta t)} \,, \nnn \\
\beta_\vk & = & \sqrt{\frac{\omega_{out}}{\omega_{in}}} \, \frac{\Gamma (1-i\omega_{in}\delta t)\Gamma(i\omega_{out}\delta t)}{\Gamma(i\omega_-\delta t)\Gamma(1+i\omega_-\delta t)}.
\label{bogocoeff}
\eea
The Heisenberg-picture state of the system is the ``in" vacuum,
\ben
a_\vk |in,0\rangle = 0 \,.
\een
We will be interested in analysing several quantities: (i) the two-point correlator of the field at a finite spatial separation (ii) the expectation value of the composite operator $\phi^2$ and (iii) the net energy density produced. In fact, the rate of change of the energy density is related to $\vev{\phi^2}$ by the Ward identity
\beq
\partial_t \vev{{\cal{E}}} = \frac{1}{2} \partial_t m^2(t) \vev{\phi^2} .
\label{ward_identity}
\eeq
%Depending on whether we want to write it as a function of the {\it{in}} or the {\it{out}} modes, 
The two point correlation function of the scalar field under the quench reads
\bea
&&\langle in, 0|\phi (\vx,t) \phi(\vx^\prime,t^\prime)|in, 0\rangle  =  \int \frac{d^{d-1}k}{(2\pi)^{d-1}}\, u_\vk (\vx,t) \,u^\star_\vk (\vx^\prime,t^\prime) 
\label{correlator} \\
& &\qquad\qquad\qquad=  \int \frac{d^{d-1}k}{(2\pi)^{d-1}} \Big\{ |\alpha_\vk|^2 ~v_\vk(\vx,t) v^\star_\vk (\vx^\prime,t^\prime)+ \alpha_\vk \beta^\star_{\vk}~v_\vk(\vx,t) v_{-\vk} (\vx^\prime,t^\prime)+ \nnn \\
& &\qquad\qquad\qquad\qquad\qquad\qquad\quad
\alpha^\star_\vk \beta_\vk ~ v^\star_{-\vk}(\vx,t)v^\star_\vk(\vx^\prime,t^\prime) +|\beta_\vk|^2 ~ v^\star_{-\vk}(\vx,t)v_{-\vk}(\vx^\prime,t^\prime) \Big\}. \nnn
\eea

We will be interested in the relationship of the results of a smooth quench as in (\ref{massprofile}) to that of an instantaneous quench from a mass $m$ to zero mass,
\ben
m_{instant}(t) = m \, \theta(-t).
\een
The ``in" and ``out" modes for such an instantaneous quench have a trivial plane wave profile for before and after the quench, respectively
\bea
u_\vk^{instant} & = & \frac{1}{\sqrt{2\omega_{in}}}e^{i(\vk \cdot \vx - \omega_{in}t)} \, , \,\,\, t\leq0 \,, \nnn \\
v_\vk^{instant} & = & \frac{1}{\sqrt{2\omega_{out}}}e^{i(\vk \cdot \vx - \omega_{out}t)} \,, \, t\geq0 \,.
\label{abruptmodes}
\eea
The Bogoliubov coefficients for the instantaneous quench are determined by demanding that the mode functions and their first derivatives are continuous at $t=0$. This yields:
\ben
\alpha_\vk^{instant} = \frac{\omega_+}{\sqrt{\omega_{in}\omega_{out}}} \qquad {\rm and} \qquad    \beta_\vk^{instant} = \frac{\omega_-}{\sqrt{\omega_{in}\omega_{out}}}.
\label{bogocoeff2}
\een
The correlator for an instantaneous quench can be easily computed using (\ref{bogocoeff2}) (or by directly matching the operator solutions across $t=0$ as in \cite{cc3}), 
\begin{eqnarray}
&&\langle in, 0|\phi (\vx,t) \phi(\vx^\prime,t^\prime)|in,0 \rangle  \rightarrow 
\int \frac{d^{d-1}k}{(2\pi)^{d-1}}\, e^{i\vk\cdot (\vx -\vx^\prime)}  \label{xprime} \\
& & \qquad\qquad\times \left[
\frac{e^{-i\omega_{out}(t-t^\prime)}}{2\omega_{out}} + \frac{(\omega_{out}-\omega_{in})^2}{4 \omega_{out}^2\omega_{in}} \cos \omega_{out}(t-t^\prime) + \frac{(\omega_{out}^2 - \omega_{in}^2)}{4 \omega_{out}^2\omega_{in}} \cos \omega_{out}(t+t^\prime) \right]. \nnn
\end{eqnarray}
The comparison of the instantaneous quench with the smooth fast quench is only meaningful at late times when the variation of the mass in the smooth quench is over, \ie  when $t\gg \dt$ and $t' \gg \dt$. For such times, the mode functions $v_\vk (\vx,t) \rightarrow v_\vk^{instant}$ which are exactly the mode functions for $t > 0$ in instantaneous quench. 

In what follows, it is useful to look at the behavior
of the Bogoliubov coefficients in various regimes.

\begin{enumerate}

\item{} First consider the limit
\ben
\omega_{in} \dt, \, \omega_{out} \dt \ll 1 \, .
\label{2-1}
\een
It may be easily checked that in this limit, the smooth quench Bogoliubov coefficients (\ref{bogocoeff}) reduce to the instantaneous quench coefficients (\ref{bogocoeff2}), regardless of the value of $|\vk|/m$. This means that the smooth quench approaches an instantaneous quench when $\dt$ is {\em small compared to all other length scales in the problem}. In particular, this means that the momentum space correlators at some momentum $\vk$ will approach the instantaneous correlator only when $m\dt \ll 1$ as well as $|\vk|\dt \ll 1$. In \cite{dgm1} we discussed the implications of this for expectation values of {\em local} quantities like $\langle\phi^2 (x,t)\rangle$ or the energy density. Generically, the small $m\dt \ll1$ limit of these local quantities will {\it{not}} agree with the instantaneous quench result since these quantities involve an integration over momenta all way upto the cutoff, and the physical smooth quenches, in which we are interested, are fast compared to physical mass scales but slow compared to the cutoff scale. It can be the case, however, that the integrand is a rapidly decaying function. If so, even if we have to integrate to arbitrarily high momenta, the main contributions will come from low momenta and then there would be agreement between the two protocols. For example, as shown in \cite{dgm2}, this is what happens in evaluating $\vev{\phi^2}$ in $d=3$. In section \ref{late2}, we will go back to this discussion and show that in higher dimensions this is not generally true.

\item{} Now consider the limit
\ben
m\dt \ll1,~~~~~|\vk| / m \gg 1,~~~~~|\vk|\dt = {\rm arbitrary} \,.
\label{2-1-1}
\een
Once again in this limit, the Bogololiubov coefficients (\ref{bogocoeff}) approach the instantaneous quench coefficients (\ref{bogocoeff2}), for any finite value of $|\vk| \dt$. In fact, in this limit the coefficients (\ref{bogocoeff}) behave as
\bea
\alpha_\vk & \rightarrow & 1 -\frac{m^2}{4k^2} [1-i k \dt \, \psi(1-ik\dt)+ik\dt \, \psi(-ik\dt) ] + O(m^4/k^4) \,, \nnn \\
\beta_\vk & \rightarrow & \frac{m^2}{ik^2}k\dt \, \Gamma(1-ik\dt) \, \Gamma(ik\dt)+ O(m^4/k^4) \,,
\label{2-2}
\eea
where $\psi(x)$ denotes the digamma function, \ie $\psi(x)=\partial_x \log \Gamma(x)$. For the instantaneous quench, instead, they behave as
\bea
\alpha_\vk^{instant} & \rightarrow & 1 + O(m^4/k^4) \,, \nnn \\
\beta_\vk^{instant} & \rightarrow &-\frac{m^2}{4k^2}\,.
\label{2-3}
\eea
Thus to leading order in $m^2/k^2$, we have $\alpha_\vk = 1, \beta_\vk
= 0$ for both smooth and instantaneous quenches regardless of the value of $k\dt$. This is a reflection of the fact that very high momentum modes
are not excited by the quench, \ie to leading order the quench is
immaterial for these modes. 
The subleading terms in (\ref{2-2}) are
of course dependent on $\dt$. In fact, for finite $k\dt$, the subleading
term in $\alpha_\vk$ is $O(m^2/k^2)$. However, if in addition we have
$k \dt \ll 1$, this $O(m^2/k^2)$ term is cancelled, as it should be.

\end{enumerate}

In the next section we discuss the implications of these observations
for real space correlation functions.

%Finally, the one important fact that we found in \cite{dgm} is that if we take the limit of long times,  but also (and very important) the limit of low energies, \ie $\omega \dt \ll 1$ for every $\omega$ in the problem, we recover the instantaneous quench correlator \cite{cc3}

%In \cite{dgm} we discussed about the implications and validity of taking the limit of low energies. In general, it is not a valid assumption to make, since in order to obtain any physical quantity we need to integrate over all momentum modes. So even if for low momenta the limit seems reasonable, at some point of the integration momentum would become large enough that the starting assumption would break down. It can be the case, however, that the integrand is a rapidly decaying function. If so, even if we have to integrate to arbitrarily high momenta, all the main contributions will come from low momenta and then our starting assumption will be reasonable. In fact, as shown in \cite{dgm}, this is what happens for $\vev{\phi^2}$ in $d=3$. In section \ref{late2}, we will go back to this discussion and show that in higher dimensions this is not generally true.
 
\section{Late time spatial correlators} \label{late} 

%Our aim here is to explore the expectation stated above, namely, that at times long after the quench and for correlators involving large separations, there should not be
%much difference between the instantaneous and the smooth fast
%quenches. 

In this section, we examine equal time correlation functions at finite spatial separations
\beq
C(t, \vec{r}\,) \equiv \langle \phi(t, \vec{r}\,) \phi(t, \vec{0}) \rangle\,, \label{coresample}
\eeq
and compare the result for smooth fast quenches and instantaneous
quenches at late times. 
%To simplify the problem, we have assumed here
%that both operators are at same time $t$. 
For simplicity, we only explicitly consider the correlators in odd
spacetime dimensions in the following. 
%Some basic results for spatial correlators of a scalar field with a constant mass are sketched in Appendix \ref{const_mass_corr}, that will help us understand the quench correlators.
We will consider this correlator in eq. (\ref{coresample}) in three different situations: the first one is the
equal time correlator for a smooth quench from a mass $m$ to zero mass, as in eq. (\ref{massprofile}),
\bea
C(t, \vec{r})_{smooth}  \label{full_late2}
= \frac{1}{2 (2\pi)^{d-1}}\int d^{d-1}k \, \, \frac{e^{i \vec{k} \cdot \vec{r}}}{k} \, \, \Big\{ |\alpha_\vk|^2+|\beta_\vk|^2  + \alpha_\vk \beta^\star_{\vk}~e^{2 i k t}+
\alpha^\star_\vk \beta_\vk ~e^{-2 i k t} \Big\}, \nnn \\
\eea
where $\alpha_\vk$ and $\beta_\vk$ are given by eq.~(\ref{bogocoeff}).
The second quantity is the equal time correlator for an instantaneous quench which can be read off from eq.~(\ref{xprime}),
\beq
C(t, \vec{r})_{instant}  = \frac{1}{2 (2\pi)^{d-1}}\int d^{d-1}k \, \, e^{i \vec{k} \cdot \vec{r}} \left( \frac{1}{ k^2 \sqrt{k^2 + m^2}} \left( k^2 + m^2 \sin^2(kt) \right) \right) \, . \label{longtime_full_corr}
\eeq
This correlator was studied in \eg \cite{cc3}. Finally, we consider the correlator
for a constant mass $m=0$,
\ben
C_{const}(\vec{r}) =  \frac{1}{2 (2\pi)^{d-1}}\int d^{d-1}k \, \, e^{i
  \vec{k} \cdot \vec{r}}\frac{1}{|\vk|}.
\label{mzerocorrelator}
\een
Constant mass correlators are evaluated in detail in Appendix \ref{const_mass_corr}, including the case of $m=0$ --- see eq. (\ref{nomass}). Performing the angular integrals above we find 
\bea
C(t, \vec{r})_{smooth} 
& = & \frac1{\sigma_c} \int {\cal{C}}_{smooth}(k,t,r)\, dk \label{baggy1} \\
& = & \frac1{\sigma_c\,r^{\frac{d-3}{2}}} \int dk \, \,  k^{\frac{d-3}{2}} \,J_{\frac{d-3}{2}}(k r) \, \Big\{ |\alpha_\vk|^2 +|\beta_\vk|^2+ \alpha_\vk \beta^\star_{\vk}~e^{2 i k t}+
\alpha^\star_\vk \beta_\vk ~e^{-2 i k t}  \Big\}, \nnn \\
C(t, \vec{r})_{instant} 
& = & \frac1{\sigma_c} \int {\cal{C}}_{instant}(k,t,r)\, dk \label{baggy2} \\
& = & \frac1{\sigma_c\,r^{\frac{d-3}{2}}} \int dk \, \, k^{\frac{d-3}{2}} \,J_{\frac{d-3}{2}}(k r)
\, \frac{k^2 + m^2 \sin^2(kt) }{ k\sqrt{k^2 + m^2}}\,, \nnn \\
C_{const}(\vec{r}) & = & \frac1{\sigma_c} \int
{\cal{C}}_{const}(k,r)\, dk = \frac1{\sigma_c\,r^{\frac{d-3}{2}}} \int
dk\,\,k^{\frac{d-3}{2}}\, J_{\frac{d-3}{2}}(k r)\,, \label{const_mass_2}
\eea
where $\sigma_c=2^{\frac{d+1}{2}}\pi^{\frac{d-1}{2}}$ and
$J_{\frac{d-3}{2}}$ is the Bessel function of order $\frac{d-3}{2}$.

Before proceeding with the detailed calculations, let us present our intuitive expectations for the comparison, as well as the results found below.  
As in section \ref{responseq}, we are considering quenches which take the mass from some fixed initial value $m$ to zero after the quench. There are several different dimensionful parameters at play in our correlators, \ie $t$, $\dt$, $r=|\vec{r}\,|$ and $m$, and so first, we wish to clearly specify how the corresponding scales are ordered in our considerations below. First, for the smooth quenches, we are considering the fast quench limit and hence we have $m\dt\ll1$. We are also examining the correlators in late time limit and hence $\dt\ll t$. While these inequalities do not fix a relation between $m$ and $t$, we will only present results for the case $mt>1$. That is, the following discussion explicitly considers quenches where 
\beq
\dt\ll1/m<t\,. \labell{order}
\eeq
We have verified that the general behaviour is the same in regimes where $mt \lesssim 1$, as long as the inequalities for the fast quench  and late time limits are both satisfied. 

Given the ordering in eq.~\reef{order}, we still have one remaining scale unspecified, namely the spatial separation $r=|\vec{r}\,|$. In the following, we compare the correlator \reef{coresample} for fast smooth quenches and that (\ref{longtime_full_corr}) for instantaneous quenches for $r$ in different regimes. The natural intuition is that in the Fourier transform, only momenta satisfying $k\lesssim 1/r$ will contribute significantly to the correlator.  Hence given that we are in the fast quench regime with $m\dt\ll1$, if we further choose $r\gg \dt$, then both of the inequalities in eq.~\reef{2-1} should be satisfied for the momenta contributing to the correlators. Hence the analysis in the previous section would indicate that at late times, the integrand in eq.~\reef{correlator} for the smooth quenches effectively reduces to that in eq.~\reef{xprime} matching the correlator for an instantaneous quench. We will explicitly verify that this expectation is realized by numerically comparing the full expression (\ref{correlator}) for smooth quenches with the instantaneous quench result (\ref{xprime}) at late times. 

Continuing with the above intuition, differences
between the late-time correlators for the two different quench
protocols are expected to arise in a regime where the
spatial separation is comparable to the quench time, \ie
$r\lesssim\dt$. In view of eq.~\reef{order}, this means that we would
be examining the correlator at short distance scales. However, we also
found that for very large $k \gg m$ the leading behavior of the
Bogoliubov coefficients are in fact independent of the quench
rate --- see discussion following eq. (\ref{2-1-1}). This immediately implies that the leading singularity at small
$r$ is independent of any quench protocol and therefore
one gets the same singular behaviour as the constant mass correlators
in this regime, namely, $C(t, \vec{r}\,) \propto 1/r^{d-2}$ --- see
Appendix \ref{const_mass_corr}. Hence the leading behaviour in the
correlators produced by the smooth and instantaneous quenches again
agrees in this regime. As we will show below, the subleading
singularities are in fact different in high dimensions.

The interesting regime where the difference between a smooth and
instantaneous quench would show up is therefore the intermediate values of
$r$. To study this difference it is convenient to eliminate the
leading small-$r$ contribution by calculating the difference between
the two quench correlators, \ie $C_{smooth}(t,
\vec{r}\,)-C_{instant}(t, \vec{r}\,)$, or alternatively by
subtracting the fixed mass correlator from each of the quench
correlators individually. 
As we describe in detail below, this difference of the late-time correlators indicates that the subleading behaviour, in fact, also agrees for $d=3$ but a small finite difference arises for $d=5$. For $d=7$ and higher dimensions, the difference still diverges in the limit $r\to0$.

\subsection{Numerical results for various dimensions}

We now evaluate the $k$ integrals in eqs. (\ref{baggy1}) -
(\ref{const_mass_2}) numerically.  However the integrands are
typically oscillating rapidly with a growing envelope. Hence to
evaluate the integral properly, we need to regulate the integral. We do so by introducing a convergence
factor $\exp (-ak)$ with $a > 0$ and evaluating the integral in the limit $a\to0$, as discussed in Appendix
\ref{const_mass_corr}. Using this regulator, the integral in
eq. (\ref{const_mass_2}) can be evaluated to yield
$C_{const}(\vec{r})\propto 1/r^{d-2}$, as in eq.~\reef{nomass}. This
is the standard massless propagator in $d$ dimensions. For our numerical calculations, we typically use $a=10^{-3}$, which we can verify is small enough that the integrals accurately converge to their limiting values. 

%To begin, we observe that long after the quench, the spatial correlators are in fact very similar to the constant mass correlator (with $m=0$), independently of whether it was an instantaneous or a smooth quench. The constant (zero) mass correlator is derived in Appendix \ref{const_mass_corr} and is given by

%Of course, this last integral can be evaluated to yield to yield $C_{fixed}(\vec{r})\propto 1/r^{d-2}$, as in eq.~\reef{nomass}. However, our present purpose is to compare this correlator with eqs.~\reef{baggy1} and \reef{baggy2} and hence we leave this expression in its integral form.

%One can easily verify that the three integrands are very similar by
%plotting them as a function of $k$ for different values of $r, t$ and
%$\dt$, \eg see fig.~\ref{fig_corr_int}. This agreement can also be
%seen by examining eqs.~(\ref{baggy1} -- \ref{const_mass_2})
%directly.\footnote{Note that with $k\gg m$, one finds from
 % eq.~(\ref{bogocoeff}) that $|\alpha_\vk|^2\simeq 1$ and
 % $|\beta_\vk|^2, |\alpha_\vk \beta^\star_{\vk}|\ll 1$.} 

\begin{figure}[h!]
\setlength{\abovecaptionskip}{0 pt}
\centering
\includegraphics[scale=0.8]{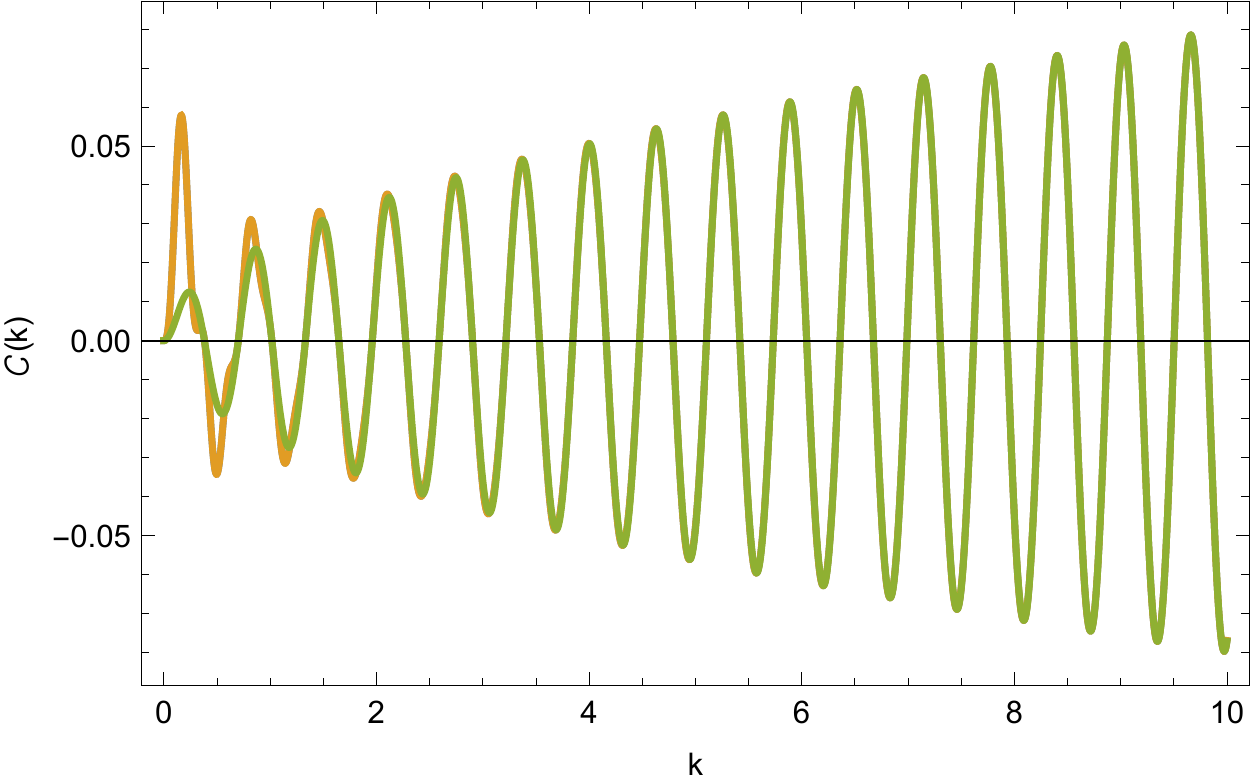}
\caption{(Colour online) Integrands ${\cal{C}}_{instant}$ (blue), ${\cal{C}}_{smooth}$ (yellow) and ${\cal{C}}_{const}$ (green) of the instantaneous quench, smooth quench and constant massless case, for $d=5$, $t=10$, $r=10$ and $\dt=1/20$ as a function of momentum $k$ --- all dimensionful quantities are given in units of the initial mass $m$. All three integrands are nearly identical for large momenta. Further the instantaneous and smooth quench curves overlap at all scales.} \label{fig_corr_int}
\end{figure}

As noted in the above discussion, for large $k \gg m$, ${\cal{C}}_{smooth}(k)$ and
${\cal{C}}_{instant}(k)$ are essentially identical to
${\cal{C}}_{const}(k)$, implying that the leading divergence in all
the corresponding correlators is  $1/r^{d-2}$ for small $r$. Hence, the
integrands are very close to each other for a large range of
$k$, as illustrated in fig.~\ref{fig_corr_int}.
%Hence we should expect that the quench correlators only differ significantly from the constant mass correlator \reef{const_mass_2} at large separations, \ie $r\gtrsim 1/m$. In particular then, for either of the quench protocols, the correlators still diverge as $1/r^{d-2}$ as $r\to0$.
%

%Hence we see that to a large extent, the quench correlators match the
%(constant) massless correlator \reef{const_mass_2}. 

Therefore, in order to highlight the differences between the smooth and instantaneous quenches, we will subtract off eq.~\reef{const_mass_2} to define for both cases
\beq
\tilde{C}(t,r) = \frac1{\sigma_c} \int  \tilde{{\cal{C}}}(k,t,r)\,dk \equiv C(t,r) - C_{const}(r) = \frac1{\sigma_c} \int \left( {\cal{C}}(k,t,r)-{\cal{C}}_{const}(k,t,r) \right) dk\,.
\label{harvest}
\eeq
In terms of the integrands, subtracting ${\cal{C}}_{const}(k)$ suppresses the growing oscillations at large $k$ that, \eg we see in fig.~\ref{fig_corr_int}.  In the full correlator, this subtraction removes the divergent behaviour at $r\to0$, which makes it easier to identify differences in the finite remainder.  If this behaviour was not removed, for instance, both the instantaneous and the smooth quench would both grow as $1/r^{d-2}$ in exactly the same way as $r\to0$ and it would be extremely difficult to find any differences between the correlators for the two different quench protocols in this regime.\footnote{This would be analogous to analysing the bare expectation value $\vev{\phi^2}$ instead of the renormalized quantity $\vev{\phi^2}_{ren}$ in \cite{dgm1,dgm2}. Of course, the interesting physical quantity is in the renormalized expectation value --- see further discussion in section \ref{smallrct}.} In comparing the subtracted integrands $\tilde{{\cal{C}}}$ below, we start by considering $d=5$. 

\begin{figure}[H]
        \centering
        \subfigure[$\dt=1/20$]{
                \includegraphics[scale=0.55]{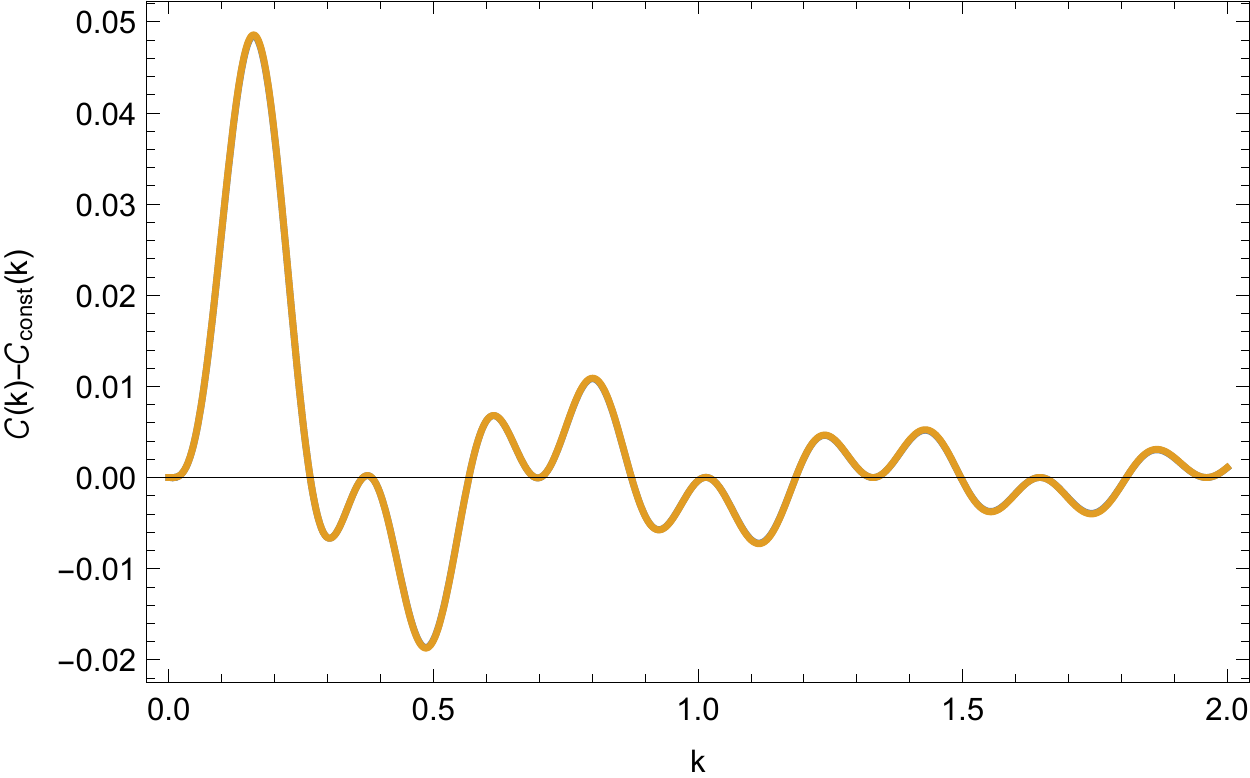} \label{fig_small_k_2}}
        \subfigure[$\dt=1/2$]{
                \includegraphics[scale=0.55]{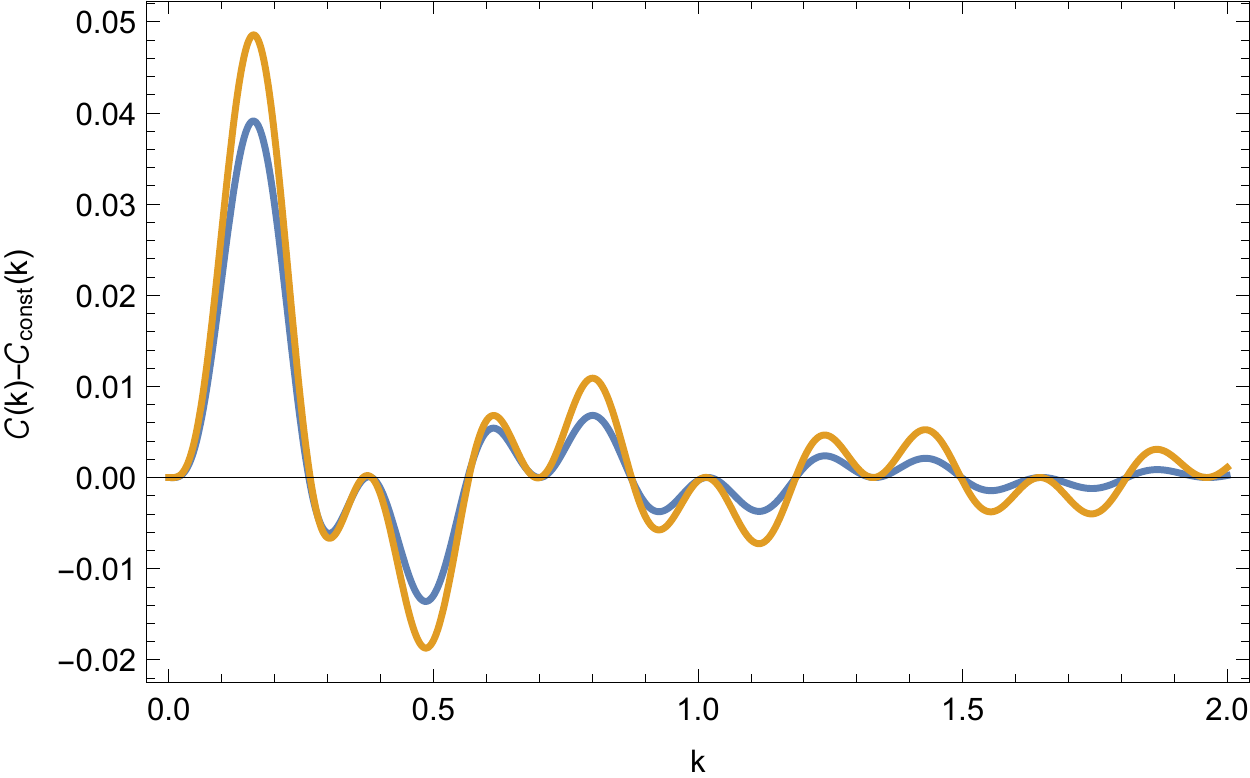} \label{fig_small_k_1}}
        \caption{(Colour online) Subtracted integrands as a function of (small) momentum $k$. In this case we are plotting for $d=5$, $t=10$, $r=10$ (with the units set by $m$). The yellow line corresponds to the instantaneous quench while the blue one to the smooth. Panel (a) shows that no detectable differences appear with $\dt=1/20$ in the range $0\le k\le2$. However, in panel (b), minor differences occur in this range when $\dt=1/2$.}
                \label{fig_small_k}
\end{figure}

In fig.~\ref{fig_corr_int}, as well as the good agreement between the quench correlators and the constant massless correlator at large $k$, we see that the two integrands, ${\cal{C}}_{smooth}(k)$ and ${\cal{C}}_{instant}(k)$, agree for throughout the momentum range shown there. Given our discussion at the opening of this section, we only expect differences to arise when $k\gtrsim 1/\dt$. One way to illustrate these differences is to make $\dt$ larger, as illustrated in fig.~\ref{fig_small_k}. Panel (a) shows the subtracted integrands in the range $0\le k\leq 2$ with the same parameters as used in  fig.~\ref{fig_corr_int}, \ie $mt=10$, $mr=10$ and $m\dt=1/20$ for $d=5$, and the two curves precisely overlap in this momentum range. The only change of parameters in panel (b) is that here $m\dt=1/2$ and we clearly see that small differences appear between the integrand for the smooth quench and that for the instantaneous quench. Note, however, that with $m\dt=1/2$, the smooth quench only barely satisfies eq.~\reef{2-1} and so while useful to illustrate possible differences, this example is not really in the fast quench regime.

\begin{figure}[H]
        \centering
        \subfigure[Intermediate momentum]{
                \includegraphics[scale=0.55]{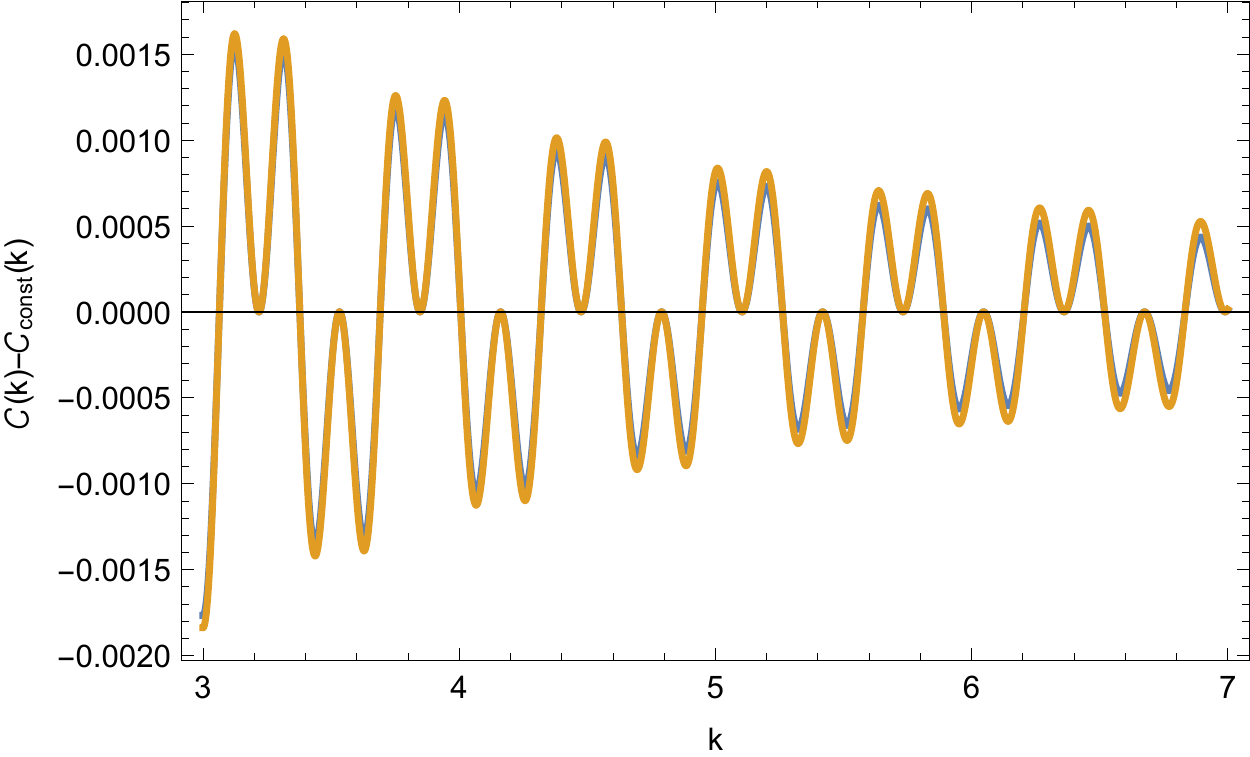} \label{fig_int_k}}
   		 \subfigure[Large momentum]{
                \includegraphics[scale=0.55]{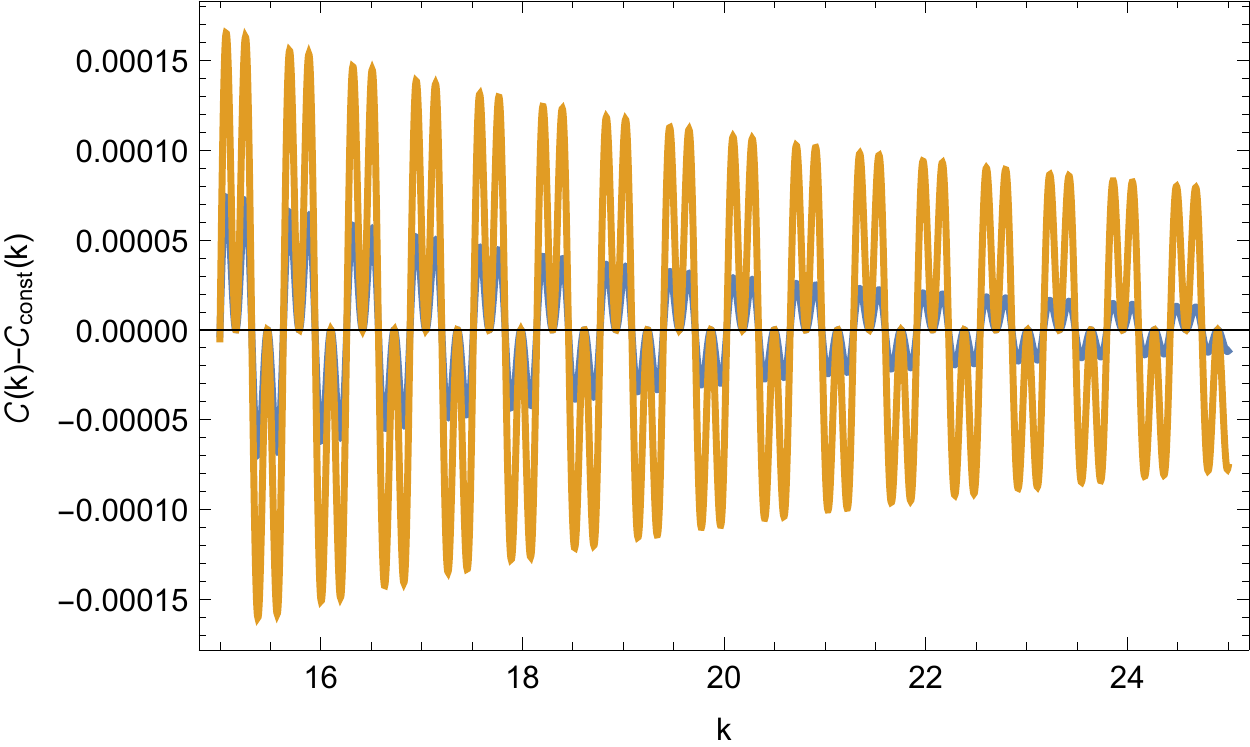} \label{fig_large_k_1}}
        \caption{(Colour online) Subtracted integrands as a function of momentum $k$. In this case we are plotting for $d=5$, $t=10$, $r=10$ and $\dt=1/20$ (with the units set by $m$). The blue line corresponds to the smooth quench while the yellow one to the instantaneous quench. Panel (a) shows an intermediate regime where no significant differences between the two integrands are apparent. Visible differences appear for larger $k\gtrsim1/\delta t$, in panel (b).}
                \label{fig_large_k}
\end{figure}

Focusing on the parameters $mt=10$, $mr=10$ and $m\dt=1/20$ (for
$d=5$), we see in fig.~\ref{fig_large_k} that the subtracted
integrands continue to agree for much larger values of $k$. However,
with $k/m \sim 20$, differences are clearly visible in panel (b).
However, comparing the vertical scale in these two plots to that in
panel (a) of fig.~\ref{fig_corr_int}, we see that these large $k$
contributions to the {\it subtracted} integrand are highly
suppressed.\footnote{Further, these large $k$ contributions are also
  highly oscillatory, so they will tend to cancel out in the
  integral.} Hence the differences should not be expected to
contribute to the full integral, \ie they should not produce significant differences in the position-space correlators.
\begin{figure}[H]
        \centering
        \subfigure[$mr=1$]{
                \includegraphics[scale=0.55]{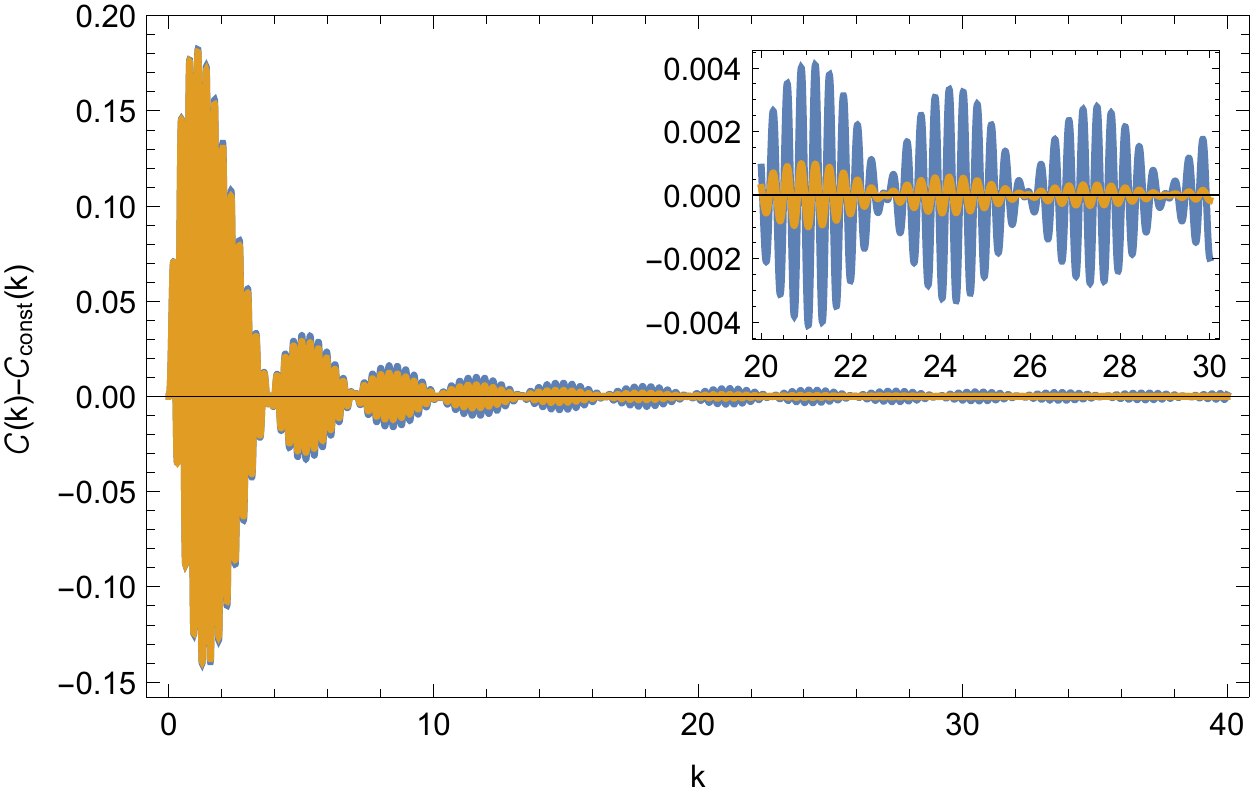}}
   		 \subfigure[$mr=0.1$]{
                \includegraphics[scale=0.55]{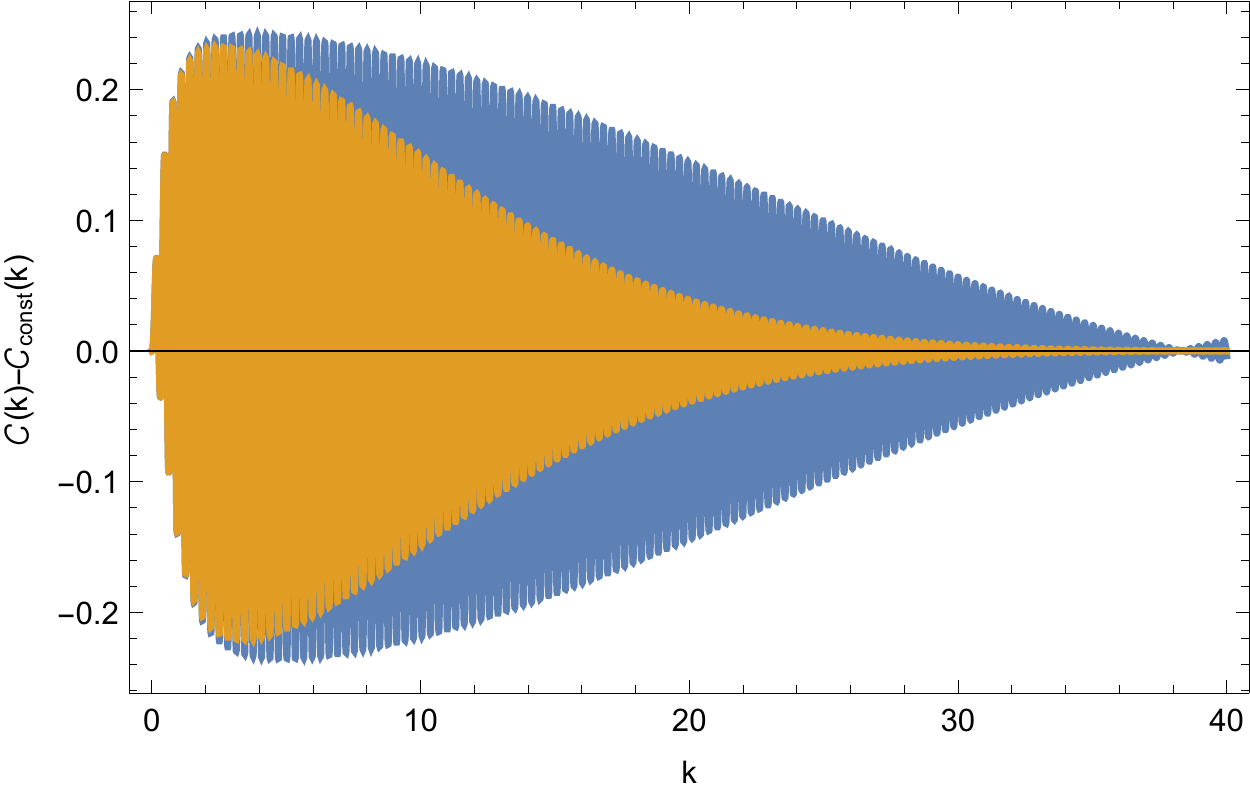}}
        \caption{(Colour online) Regulated integrands as a function of momentum $k$ for $d=5$ and small separation. The blue line corresponds to the smooth quench while the yellow one to the instantaneous. At small distances the two integrands are clearly different from each other, even at scales $k \dt \leq 1$.}
                \label{fig_d7_corr}
\end{figure}

It turns out that examining the subtracted integrands for $d=3$ yields essentially the same results. On the other hand, the situation is also similar in $d=7$ for long distances. As we decrease $r$, for $d=7$, we see analogous effects to those in fig.~\ref{fig_d7_corr}, \ie the two integrands become clearly different. However, as we will see, after integration, the behaviour of the correlator at small distances is very different depending on the dimension.

\begin{figure}[h!]
\setlength{\abovecaptionskip}{0 pt}
\centering
\includegraphics[scale=1]{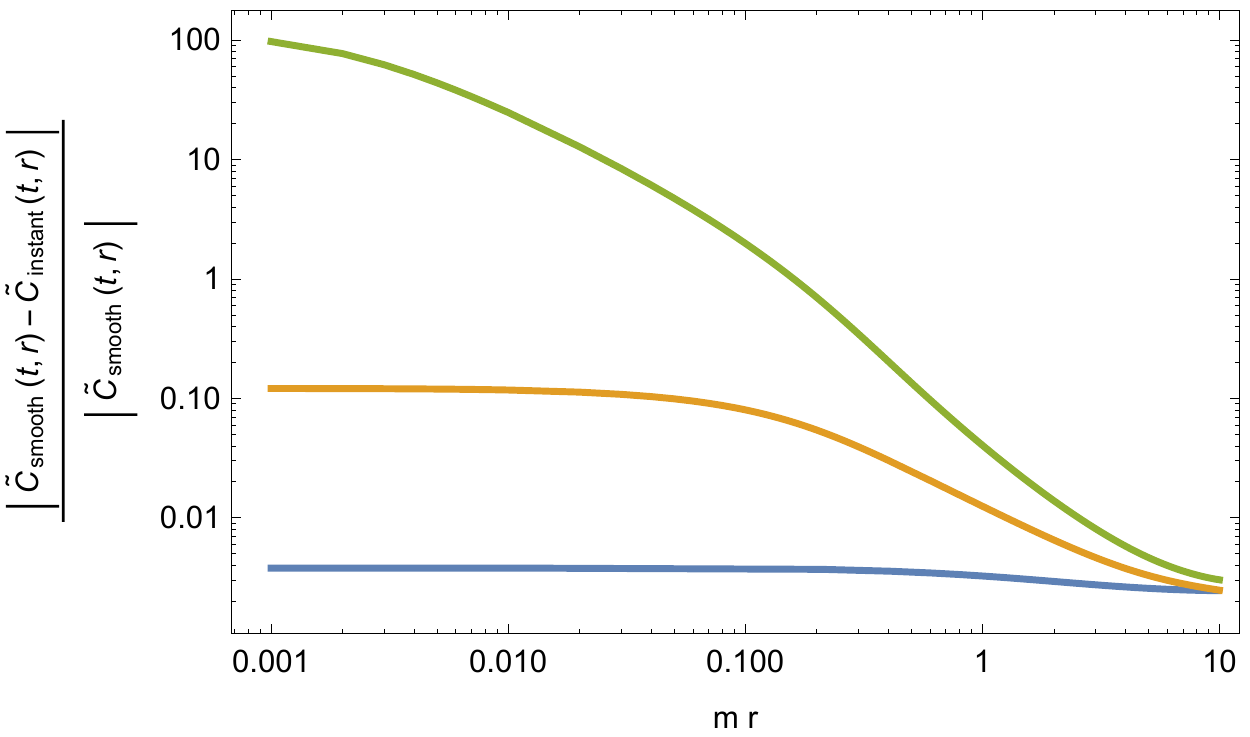}
\caption{(Colour online) Difference between the late-time correlators for smooth and instantaneous quenches as a function of the separation distance $r$. The blue line corresponds to the $d=3$ case while the yellow one belongs to $d=5$ and $d=7$ is shown in green. We are using $m t=10$ with $m \dt=1/20$. In $d=3$ and $d=5$, the difference remains small for any value of $r$, while in $d=7$, it seems to diverge as $r\to0$. \label{fig_dist_corr}}
\end{figure}
Given that we understand the behaviour of the (subtracted) integrands, let us now compare the (position-space) correlators for instantaneous and smooth quenches at different values of $r$, as shown in fig.~\ref{fig_dist_corr}. In the figure, we have chosen $mt=10$ and $m\dt=1/20$ while $mr$ varies from 10 to 0.001. 
%Note that we are normalizing the difference with $|C(t, {r})_{smooth} - C(t, {r})_{const}|$ to eliminate the divergent behaviour as $r\to0$ --- naturally, this divergence already cancels in the difference $|C(t, {r})_{smooth} - C(t, {r})_{instant}|$ appearing in the numerator.
 As we have expected, the figure shows that the difference between the correlators goes to zero, or at least is of order $O(\dt^2)$, at large $mr$.  The results for $d=3$ indicate that the difference remains vanishingly small for all values of $r$. 
In the case of $d=5$, a small but finite difference appears to develop as $r$ goes to zero. The differences in fig.~\ref{fig_dist_corr} are most evident for $d=7$. In this case, the relative difference is already of order one at $mr=0.1$ and the trend in the figure is that it continues to grow at smaller $r$. Our expectation is that in fact, this difference will in fact diverge as $r\to0$. This conclusion comes from comparing to the late-time behaviour of expectation value $\langle \phi^2 \rangle_{ren}$ in the next section.

In summary then, we have explicitly shown that at times long after the
quench, the correlators generated by instantaneous and the smooth fast
quenches are identical at large separations. As might be expected,
differences only appear at separations of the order the quench time
$\dt$. Further these differences are small in lower dimensions, \eg
$d=3,5$, but can be quite dramatic in higher dimensions. 
Interestingly, for $d=3$ and 5, the subtracted correlator of both the instantaneous and the smooth quench (with the $m r=0$ piece subtracted out) agree to a rather high precision for any distance $m r$. As we increase $m r$ both answers become even closer.
%However, we must emphasize that the correlators all exhibit the same singular behaviour, \ie $C(t,r)\propto 1/r^{d-2}$, at small separations and so the leading behaviour of the correlators still matches in this regime.

%\comment{more words:} In the case of $d=5$, as $r$ goes to zero, the error starts increasing until it gets to a value similar to that reported in the evaluation of $\langle \phi^2 \rangle$, see section \ref{late2}. 

%This kind of divergence, though, cannot be one of the usual divergences that we include in the counterterms for $\vev{\phi^2}$ because we are actually subtracting them from the very beginning. So we can predict that they are not related to divergences appearing in the constant mass case and so, they must be some particular effect of doing an instantaneous quench. It will be interesting to understand how these extra divergences arise in the case of the instantaneous quench. In fact, as we will soon see, in $d=7$ the instantaneous quench presents extra divergences when we compute $\vev{\phi^2}_{ren}$ that cannot be cured even after regulation. This means that the instantaneous and the smooth quench give infinitely different answers when computing $\vev{\phi^2}$ and these are the same differences that we are observing as $mr\to0$ in the spatial correlator.

\subsection{Small $r$ behavior and counterterms for local operators}
\label{smallrct}

%\comment{Comparing $1/r^{d-2}$ and $k^{d-3}$ divergences. Put these
%  words elsewhere:} 

%We know that if we take the spatial correlator and start decreasing the separation distance to zero, we should be recovering $\phi^2$ and the expectation value of $\phi^2$ is diverging. In fact, we can associate the $r$ divergence of the correlator with the $k$ divergence of the expectation value.

In \cite{dgm1,dgm2}, we studied the UV divergences appearing in $\vev{\phi^2}$ with a momentum cut-off in great detail. Intuitively, one can think that the correlator at small $r$ provides a point-splitting regulator of the same quantity. Hence, the divergences in the correlator at small $r$ should be related to the UV divergences of the local quantity $\langle\phi^2\rangle$. In
this subsection, we make this connection precise.

Let's start with the constant massless correlator, (\ref{const_mass_2}). For small $r$ we can expand the Bessel function to get,
\beq
J_{\frac{d-3}{2}} (kr) = (kr)^{\frac{d-3}{2}} \left( \frac{2^{\frac{3-d}{2}}}{ \Gamma \left(\frac{d-1}{2}\right)} + O\left((kr)^2\right) \right)\,.
\label{watchtower}
\eeq
Inserting this in eq. (\ref{const_mass_2}), 
%the $r$ dependence vanishes and what is left is an 
we get an integral of $k^{d-3}$, exactly the same as the leading divergence in $\vev{\phi^2}$.

%But we can go further than that and establish a precise relation between each diverging term in the expectation value with terms proportional to inverse powers of the spatial separation (that, of course, will diverge as $r\to0$).

Let us first recall the set of counterterms that we found in \cite{dgm1,dgm2} for $\vev{\phi^2}$. There the leading divergences in momentum space were determined by performing an adiabatic expansion in time derivatives and then expanding our results for large momentum. We then found that this was the same as expanding the integrand of $\vev{\phi^2}$. So, to be more specific, in the ``in"-basis, the integrand took the form $k^{d-2} \omega_{in}^{-1} |_2F_1|^2$ and for large $k$ we found \cite{dgm2}
\bea
\frac{k^{d-2}}{\omega_{in}} |_2F_1|^2 & \simeq & k^{d-3} - \frac{k^{d-5}}{2} m^2(t) + 
\frac{k^{d-7}}{8} \left( 3m^4(t)+ \partial^2_t m^2(t)\right) \label{ict} \\
& & - \frac{k^{d-9}}{32} \left(10 m^6(t) + \partial^4_t m^2(t) + 10 m^2(t)\, \partial^2_t m^2(t)+5\partial_t m^2(t)\, \partial_t m^2(t)\right)  + \cdots \,. 
\nonumber
\eea
Above, we are giving all the terms needed to regulate the
expectation value up to $d=9$. Apart from the divergent
terms in the constant mass expectation value (that will appear as zeroth
order in the adiabatic expansion), they include terms with time derivatives of the mass profile which produce divergences in $\langle\phi^2\rangle$ for $d \geq 6$. These (perhaps
surprising) terms were carefully analysed in \cite{dgm2}. Now we can express the bare expectation value of the local operator in terms of an energy cut-off $\Lambda$, obtained by integrating the momentum integral up to a maximum value $k_{max} \simeq \Lambda$. This yields
\begin{eqnarray}
\vev{\phi^2} & \simeq &  \frac{1}{2^{d-1} \pi ^{\frac{d-1}{2}} \Gamma \left(\frac{d-1}{2}\right)} \left( \frac{\Lambda^{d-2}}{d-2} - \frac{\Lambda^{d-4}}{d-4} \frac{m^2(t)}{2} + \frac{\Lambda^{d-6}}{d-6} \frac{3m^4(t)+ \partial^2_t m^2(t)}{8} - \right.  \\
& & \left. - \frac{\Lambda^{d-8}}{d-8} \frac{10 m^6(t) + \partial^4_t m^2(t) + 10 m^2(t)\, \partial^2_t m^2(t)+5\partial_t m^2(t)\, \partial_t m^2(t)}{32} + \cdots \right) \,. \nnn
\label{barephi2}
\end{eqnarray}

%Now we can write the spatial correlator in a convenient way, see eq. (), by using the ``in"-basis. It turns out that the correlator can be written as,
Working in terms of the ``in'' modes, the smooth quench correlator can be expressed as
\bea
C(t, \vec{r}\,)_{smooth} & = & \frac{1}{\sigma_c r^{\frac{d-3}{2}}} \int dk \frac{k^{(d-1)/2}}{\omega_{in}}   J_{\frac{d-3}{2}}(k r) \left|_2F_1 \right|^2 \,, \nonumber \\
& = & \frac{1}{\sigma_c r^{\frac{d-3}{2}}} \int dk \frac{1}{k^{\frac{d-3}{2}}}  J_{\frac{d-3}{2}}(k r) \frac{k^{d-2}}{\omega_{in}}\left|_2F_1 \right|^2 \,, \label{corr_to_vev}
\eea
where in the last line we have presented the integrand in a way which allows us to make use of the above expansion in eq. (\ref{corr_to_vev}). So we have all the ingredients to take the limit from the spatial correlator to the expectation value. In particular, if we take the $r\to0$, then the Bessel function can be replaced for its zeroth order expansion shown in eq. (\ref{watchtower}). The  powers of $kr$ in eq. (\ref{watchtower}) will cancel those same powers appearing in eq. (\ref{corr_to_vev}), while the numerical factors will turn the $\sigma_c$ into a $\sigma_s$.\footnote{We remind the reader that $\sigma_s$ is a numerical constant that depends on the space-time dimension $d$ and was used in \cite{dgm1,dgm2} to normalize the expectation value of $\phi^2$. Explicitly, $\sigma_s \equiv \frac{2(2\pi)^{d-1}}{\Omega_{d-2}}= \frac{\Gamma \left(\frac{d-1}{2}\right)}{2^{d-1} \pi ^{\frac{d-1}{2}}}$, where $\Omega_{d-2}$ is the angular volume of a unit $(d-2)$-dimensional sphere.} Then we can use eq. (\ref{ict}) to expand for large $k$ and find that the correlator behaves exactly in the same way as $\vev{\phi^2}$ when $r\to0$. In fact,
\bea
C(t, \vec{r}\,) & \xrightarrow[r\to0]{} & \frac{1}{\sigma_s} \int dk \left( k^{d-3} - \frac{k^{d-5}}{2} m^2(t) + 
\frac{k^{d-7}}{8} \left( 3m^4(t)+ \partial^2_t m^2(t)\right) \right. \\
& & \left. - \frac{k^{d-9}}{32} \left(10 m^6(t) + \partial^4_t m^2(t) + 10 m^2(t)\, \partial^2_t m^2(t)+5\partial_t m^2(t)\, \partial_t m^2(t)\right)  + \cdots \right) \,. 
\nonumber
\eea

%Finally we want to discuss one more issue regarding the divergences, that would be important in section \ref{uni_scaling}. 

Of course for any finite (positive) $r$ the correlator will be UV finite. So in principle we should be able to perform the integral over momenta to obtain a series expansion in inverse powers of $r$. We now show that this small-$r$ expansion of the correlator is directly related to the large-$k$ expansion for the expectation value. Let us take expression in eq. (\ref{corr_to_vev}) and replace $k^{d-2} \omega_{in}^{-1} |_2F_1|^2$ with the expansion given in eq. (\ref{ict}). The integrand is composed of a series of terms which are constants (not functions of momentum) multiplying the Bessel function and some power of $k$. For fixed value of $r$, we can integrate that expression, using 
\beq
\int_0^{\infty } k^{\alpha } J_{\frac{d-3}{2}}(k r) \, dk = \frac{2^{\alpha }}{r^{\alpha+1} } \frac{\Gamma \left(\frac{1}{4} (d+2 \alpha -1)\right)}{\Gamma \left(\frac{1}{4} (d-2 \alpha -1)\right)}. \label{int_alpha}
\eeq
Now, as an example, consider the leading divergence, \ie $k^{d-3}$, which gives
\beq
\frac{1}{\sigma_c r^{\frac{d-3}{2}}} \int_0^{\infty } k^{\frac{d-3}{2}} J_{\frac{d-3}{2}}(k r) \, dk = \frac{1}{\sigma_c} \frac{2^{\frac{d-3}{2}} \Gamma \left(\frac{d}{2}-1\right)}{\sqrt{\pi }} \frac{1}{r^{d-2}},
\eeq
after using eq. (\ref{int_alpha}) with $\alpha = (d-3)/2$ and some algebra. This
shows that UV divergences of the local quantity $\langle\phi^2\rangle$ appear as inverse powers of $r$ in the finite spatial separation correlator, \ie $r$ plays the role of a point-splitting regulator, replacing the momentum cut-off $\Lambda$ in eq. (\ref{barephi2}). In the above example we showed that the leading divergence is proportional to $r^{d-2}$. We can proceed to do the integral for $\alpha = \frac{d-3}{2}-2$. This would correspond to  divergences proportional to $k^{d-5}$ and will lead to a term in the spatial correlator which is inversely proportional to $r^{d-4}$. Note that in general, eq. (\ref{int_alpha}) maps the integral over $k^\alpha$ to the power $1/r^{\alpha+1}$. Also note that there is an important difference between the leading divergence and the rest of the divergent terms. All of the subleading divergences are proportional either to the instantaneous mass $m(t)$ or to time derivatives of $m(t)$, while the leading divergence is independent of $m(t)$. 
%Knowing already the counterterms for the expectation value of $\phi^2$, we can then have a concrete idea on how the correlator will behave as a power series in $r$. 
This means that the leading term as $r\to0$ will be inversely proportional to $r^{d-2}$ but then there will be subleading terms inversely proportional to $r^{d-4}, r^{d-6},$ etc., that will also contain factors of the mass and its derivatives. Explicitly we find
\bea
& & C(t,r)  \simeq \frac{\Gamma \left(\frac{d-4}{2}\right)}{8 \pi ^{d/2}} \left( \frac{d-4}{r^{d-2}} -  \frac{1}{ r^{d-4}} \frac{m^2(t)}{2} + \frac{1}{3 (d-6) r^{d-6}} \frac{3m^4(t)+ \partial^2_t m^2(t)}{8} - \right. \\ 
& & \left. - \frac{1}{15 (d-6) (d-8) r^{d-8}} \frac{10 m^6(t) + \partial^4_t m^2(t) + 10 m^2(t)\, \partial^2_t m^2(t)+5\partial_t m^2(t)\, \partial_t m^2(t)}{32} + \cdots \right) \,. \nnn
\label{corr_r_exp}
\eea
In particular, note that to analyse the structure of the correlator in $d=7$, we would have to take into account a term that is proportional to the second derivative of the mass, that will increase as $1/r$ when $r\to0$. This will be important to understand the behaviour of the correlator in section \ref{uni_scaling}.
Also note that this term is also subleading compared to the term proportional to $m^2/r^{3}$, that comes second in the expansion of eq. (\ref{corr_r_exp}).

Finally let us note that, even though the momentum cut-off expression in eq. (\ref{barephi2}) and the $r$ expansion expression in eq. (\ref{corr_r_exp}) are similar in form, there is no simple way to relate the energy scale $\Lambda$ with the point-splitting regulator $1/r$. Rather, equating these two equations we get,
\beq
\Lambda \simeq \frac{\Gamma (d-1)^{\frac{1}{d-2}}}{r} \left(1+ \frac{\left(1-\Gamma (d-2) \Gamma (d-1)^{-\frac{d-4}{d-2}}\right) }{2 (d-4) \Gamma (d-1)^{\frac{2}{d-2}}} m^2(t) r^2  + \cdots \right) \,.
\eeq
This simply points-out that these divergent terms are unphysical and that these two regularization schemes have slightly different counterterms.

\section{Universal scaling in quenched spatial correlators}
\label{uni_scaling}
In \cite{dgm1,dgm2} we found an interesting set of universal scaling relations for the expectation value of the quenched operator $\langle\phi^2\rangle$ and the energy density. For the quenches considered in this work, this scaling takes the form $\vev{\phi^2}_{ren} \sim m^2 \dt^{4-d}$. We also found analytic leading order expressions for this expectation value in the case of $\dt \to 0$. For odd spacetime dimensions, we found
\beq
\langle\phi^2\rangle_{ren} = (-1)^{\frac{d-1}{2}} \frac{\pi}{2^{d-2}}\, \partial^{d-4}_t m^2(t) + O(\dt^{6-d}),
\label{phi_odd}
\eeq
which reproduces the above scaling with the mass profile \ref{massprofile} we are considering.
On the other hand, we have already shown how the spatial correlator approaches the expectation value of $\phi^2$ as the separation distance goes to zero. Then an interesting question to ask is whether there are any signs of the universal scaling in the spatial correlator.

To investigate on this question here, we will concentrate on early times, since this is the regime where scaling of the local quantities hold.
% in the middle of the quench, at $t=0$. 
Now we need a suitable object to compute. We remind the reader that the scalings hold for renormalized expectation values. Given that the bare expectation values are UV divergent, we had to add suitable counterterms to eliminate those divergences. On the contrary the spatial correlator is finite for any finite separation $r$.  However, as discussed in the previous section, the counterterms are in precise correspondence with the small $r$ expansion of the correlator.
%the relation between the leading $r$ behaviour as $r \to0$ of the correlator and the counterterms for $\phi^2$. Basically, the correlator behaves as $r^{2-d}$ and that is closely related to the main 
In particular the leading UV divergence of $\vev{\phi^2}$ goes as $\Lambda^{d-2}$, which reflects the leading small $r$ divergence of the correlator behaving as $1/r^{d-2}$. So from this perspective, the scaling behaviour is only exhibited in a higher order term, which remains finite as $r\to0$. However, we may still see the scaling in the correlator by subtracting a suitable fixed mass correlator to remove the terms which diverge in the small $r$ limit. It turns out that in order to eliminate these divergences (which are proportional to powers of $m^2$) the interesting object to compute is the difference of the spatial quenched correlator at time $t$ with the correlator at a constant value of the mass equal to the instantaneous mass at that time $t$. The latter fixed mass correlator has been computed in Appendix \ref{const_mass_corr} and one finds,
\beq
C_{fixed}(\vec{r}) =  \frac{1}{\sigma_c\,r^{\frac{d-3}{2}}} \int \frac{k^{\frac{d-1}{2}}dk}{\sqrt{k^2+m^2(t)}}\, J_{\frac{d-3}{2}}(k r)\,.
\label{mass_corr_4}
\eeq
%with the value of the mass the same as the one the mass has during the quench at $t=0$, \ie $m^2=1/2$.
In order to numerically integrate the correlator, we will go back to the ``in" basis --- see eq. (\ref{corr_to_vev})--- to obtain
\beq
C(t, \vec{r}\,) = \frac{1}{\sigma_c r^{\frac{d-3}{2}}} \int dk \frac{k^{(d-1)/2}}{\sqrt{k^2+m^2}}   J_{\frac{d-3}{2}}(k r) \left|_2F_1 \right|^2 \,.
\label{4-1}
\eeq

%Note that at $t=0$, the correlator for the instantaneous quench, see eq.(\ref{baggy2}), is exactly the same as the fixed mass correlator.

\begin{figure}[H]
        \centering
        \subfigure[$d=5$]{
                \includegraphics[scale=1]{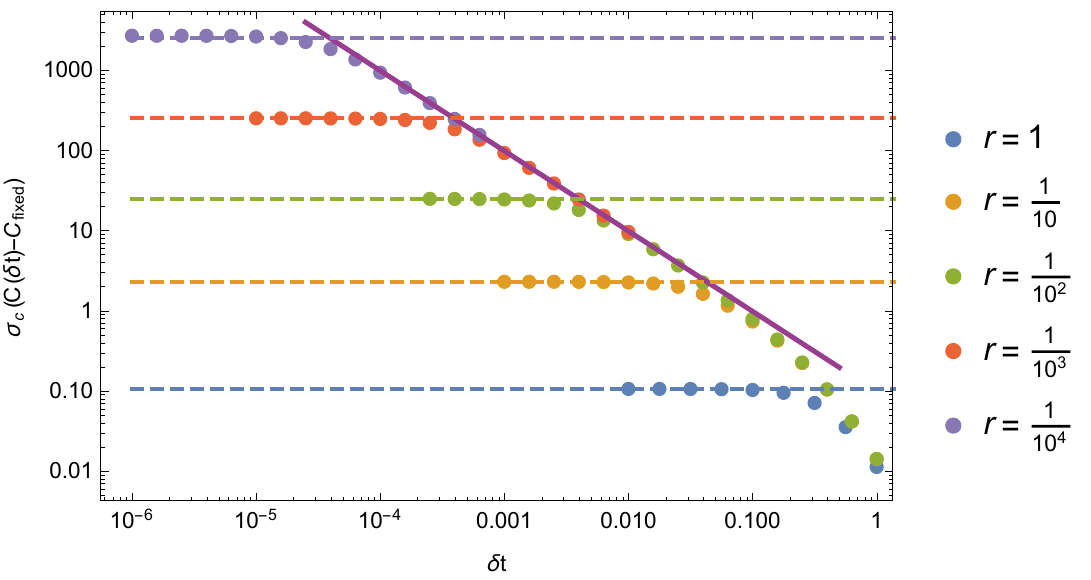} \label{d5_scaling}}
   		 \subfigure[$d=7$]{
                \includegraphics[scale=1]{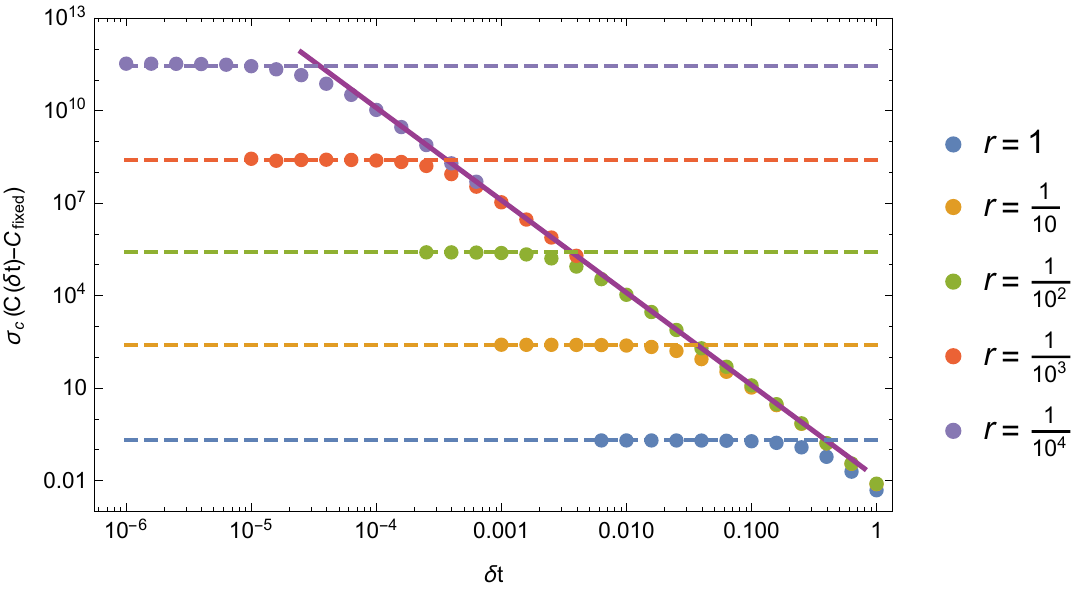} \label{d7_scaling}}
        \caption{(Colour online) Spatial correlator under a smooth quench at $t=0$ as a function of both $\dt$ and the distance separation $r$. In each case, we are subtracting the fixed mass correlator with $m^2=1/2$. The dashed lines correspond to computing the instantaneous quench correlator at $t=0$ for the different separations $r$, that is the same as computing the fixed mass correlator with $m=m_{in}=1$. The purple solid line shows the analytic leading order contribution to $\vev{\phi^2}$, given by eq. (\ref{phi_odd}).} \label{fig_scaling}
\end{figure}

First consider the correlator at $t=0$. 
In fig.~\ref{fig_scaling}, we computed $C(t=0,r)-C_{fixed}(r)$ for a wide range of values of $r$ and $\dt$ for $d=5$ and $d=7$. We see a very interesting behaviour. Basically we can divide the correlator in three different regions: (i) $r>\dt$, (ii) $r<\dt<m^{-1}$ and (iii) $r<\dt \sim m^{-1}$.

For $\dt < r$ we see that the correlator is essentially a constant independent of $\dt$ that depends on $r$. In fact it turns out that this constant value is exactly the same value of the instantaneous quench correlator evaluated at $t=0$. Recall that the instantaneous quench correlator at $t=0$ is exactly the same as the constant mass correlator with $m^2=m_{in}^2$. This coincidence might lead the reader to think that this behaviour is something special for $t=0$. However, in what follows, we will show that is the general behaviour of the correlator at any finite $t/\dt$, as long as $\dt \ll r \ll 1/m$.

Let's start by fixing the dimensionless time $\tau=t/\dt$. Now we want to analyse the $r$ and $\dt$ dependence of the following object:
\beq
C(t, \vec{r}\,)-C_{fixed}(\vec{r}) = \frac{1}{\sigma_c r^{\frac{d-3}{2}}} \int dk k^{\frac{d-1}{2}}     J_{\frac{d-3}{2}}(k r) \left( \frac{1}{\sqrt{k^2+m_{in}^2}} \left|_2F_1 \right|^2  - \frac{1}{\sqrt{k^2+m^2(\tau)}}\right) \, \,.
\eeq
Note that the first term inside the big brackets has $m^2_{in}$ in the denominator while the fixed mass part carries an $m^2$ equal to that at the particular time we are considering. 

The second important thing to notice is where is the time-dependence in the quenched correlator. The only place where $\tau$ appears explicitly is in the last argument of the hypergeometric function. Recall that
\beq
_2F_1 \equiv \, _2F_1 \left( 1+ i \omega_- \dt, i \omega_- \dt; 1 - i \omega_{in} \dt; \frac{1+\tanh(t/\dt)}{2} \right).
\eeq
Then, if we fix $\tau$ by inserting any finite number (or zero) in that last argument, we are left with an object that depends only on $\dt$ and $r$, and we can take the desired limit without any problem. So consider now $\dt \ll r$. We get that limit by considering the limit of the hypergeometric function with $\dt \to0$. As the second argument is proportional to $\dt$, to lowest order all the terms in the infinite sum will vanish but the first one, so we just get that when $\dt \to0$, $_2F_1 = 1 + O(\dt)$. Note that this argument is valid only in the case where we fix $\tau$ and effectively there is no $\dt$ dependence in the last argument. Then, after taking this first limit, we are left with
\beq
C(t, \vec{r}\,)-C_{fixed}(\vec{r}) \approx \frac{1}{\sigma_c r^{\frac{d-3}{2}}} \int dk \, k^{\frac{d-1}{2}}     J_{\frac{d-3}{2}}(k r) \left( \frac{1}{\sqrt{k^2+m_{in}^2}}  - \frac{1}{\sqrt{k^2+m^2(\tau)}}\right) \, \,.
\label{eq_corr}
\eeq
But this is nothing more than $C_{fixed}(m^2=m^2_{in}) - C_{fixed} (m^2 = m^2(\tau))$! So, at any time we get that for $\dt \ll r$ our object becomes the difference of two {\it{fixed}} mass correlators. In particular, it becomes independent of $\dt$ and that explains the horizontal dashed lines of fig.~\ref{fig_scaling}. 

However we can even go further and consider the limit of $r\ll1/m$. As it is just the limit for fixed mass correlators we can extract it directly from eq. (\ref{d7_div_corr}) in Appendix \ref{const_mass_corr}. As we are subtracting two fixed mass correlators, the leading term in the expansion, \ie the one proportional to $r^{d-2}$, will cancel and the leading contribution will come from the first subleading term:
\beq
C(t, \vec{r}\,)-C_{fixed}(\vec{r}) \approx \frac{1}{\sigma_c} \left(\frac{m^2_{in}}{2 r^{d-4}} - \frac{m^2(\tau)}{2 r^{d-4}} \right) \,.
\label{eq_corr2}
\eeq

In our plots of fig.~\ref{fig_scaling}, we have $m_{in}=1$ and $m(\tau=0)=1/2$, so the difference of correlators in the limit of $\dt \ll r \ll 1/m$ should go as $(4 \sigma_c \, r^{d-4})^{-1}$, which agrees with the values that the horizontal lines take in the plots. We can add that the agreement gets more exact as $r$ takes smaller and smaller values. 

Note, however, that eq. (\ref{eq_corr}) is valid for any value of $m$, as long as $r/\dt \gg 1$, but eq. (\ref{eq_corr2}) also needs $r m \ll 1$. If we concentrate in a regime where $r m \sim 1$ (see bottom line in each plot of fig.~\ref{fig_scaling}), then we should expect eq.(\ref{eq_corr}) to hold rather than eq.(\ref{eq_corr2}). In fact, for $r=m=1$, the bottom dashed line in fig.~\ref{fig_scaling} corresponds to approximately 0.10 and 0.20 for $d=5$ and $d=7$, respectively. This is in perfect agreement with eq. (\ref{eq_corr}) and clearly differs from the 0.25 that eq. (\ref{eq_corr2}) is predicting.

\begin{figure}[H]
        \centering
        \subfigure[$\tau = 1/4$]{
                \includegraphics[scale=1]{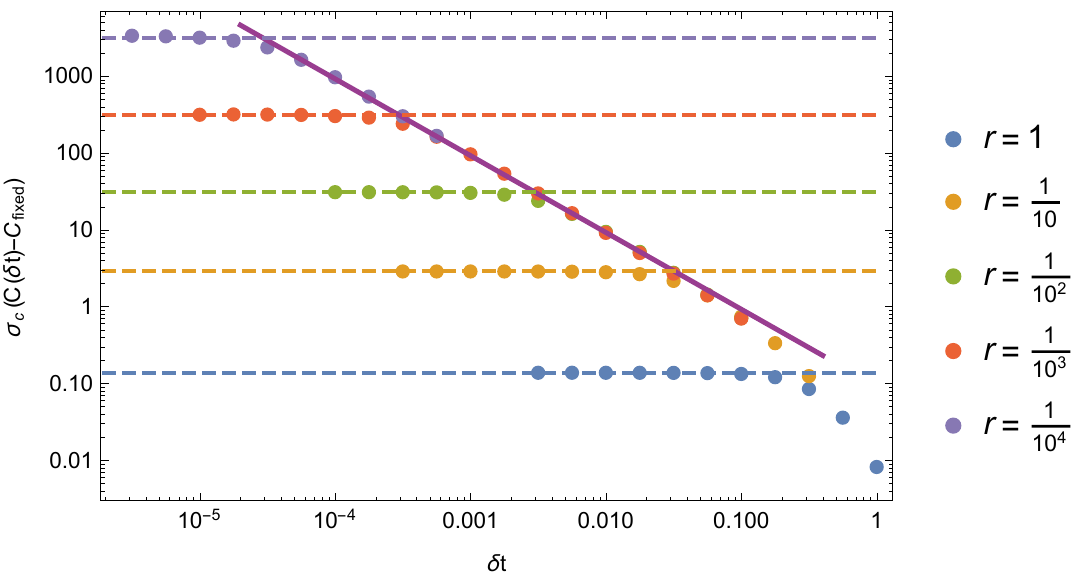} \label{t_quarter}}
   		 \subfigure[$\tau = 1/2$]{
                \includegraphics[scale=1]{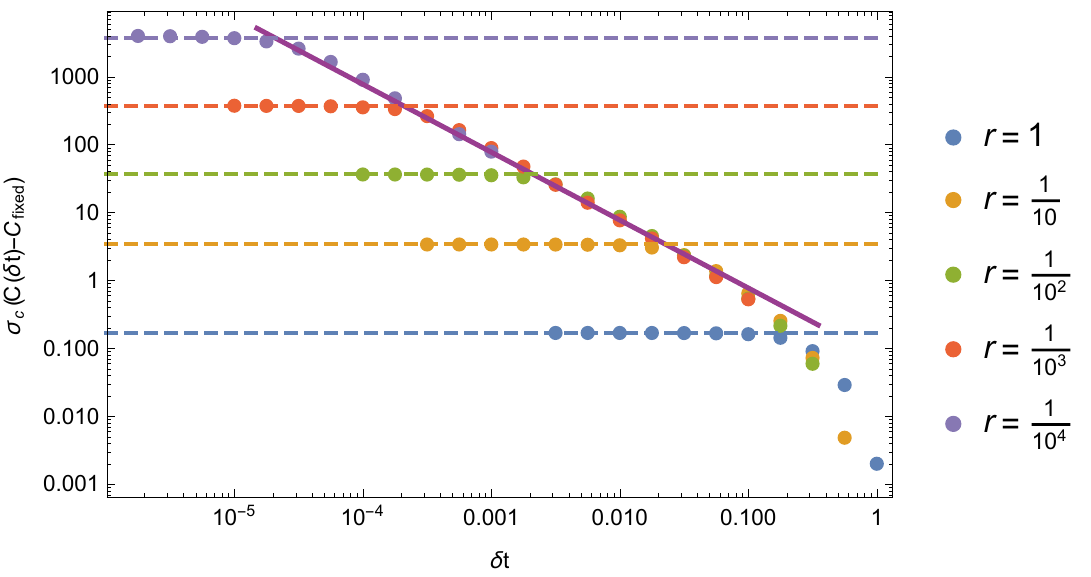} \label{t_half}}
   		         \caption{(Colour online) Spatial correlator under a smooth quench for fixed $t/\dt = \tau$ as a function of both $\dt$ and the distance separation $r$. In each case, we are subtracting the fixed mass correlator with $m^2(t/\dt=\tau)$. The dashed lines correspond to computing the fixed mass quench correlator with $m=m_{in}$ for the different separations $r$. The purple solid line shows the analytic leading order contribution to $\vev{\phi^2}$, given by eq. (\ref{phi_odd}).}
\label{fig_scaling2}
\end{figure}

To further support our claim we provide equivalent plots but at different times for $d=5$. In fig.~\ref{fig_scaling2}, we examine the results for $\tau=1/2$ and $\tau=1/4$. In these cases, the expectation is that the limiting value would be
\beq
C(t, \vec{r}\,)-C_{fixed}(\vec{r}) \approx \frac{m^2_{in} - m^2(\tau)}{\sigma_c \, r} = \frac{1-(1/2-1/2 \tanh \tau)}{\sigma_c \, r},
\eeq
and that's what arises in this figure.\footnote{For $\tau=1/2$, then $C(t, \vec{r}\,)-C_{fixed}(\vec{r}) \approx  0.365529/(\sigma_c \, r)$ and for $\tau=1/4$, the value is $0.31123/(\sigma_c \, r)$.}

\
\

The second regime $r<\dt<1/m$  leads to the most interesting behaviour. The correlator now exhibits exactly the same scaling as the $\phi^2$ expectation value. The solid lines we see in the plots are exactly the lines that come from evaluating eq. (\ref{phi_odd}) for $d=5$ and $d=7$ with the present mass quench profile --- see eq. (\ref{massprofile}). This means that in this regime exactly the same universal scaling we've been discussing for $\phi^2$ is reproduced in the spatial correlator.

Finally, at least for $d=5$, we see that when $\dt \sim 1/m$ this behaviour breaks down and our calculation goes away from the universal scaling line. It would be interesting to understand better this {\it{slow}} regime as it could be connected to other set of important universal scalings in quantum quenches: namely, the Kibble-Zurek scaling \cite{kibble, zurek} and this would give a connection between the {\it{fast}} and the {\it{slow}} regime universality in quantum quenches. We hope to report on this in the near future.

%In figure (\ref{fig_scaling2}) we show the results of these correlators at $t/\dt = 1/2$. Once again, for small $r/\dt$ there is excellent scaling, whereas for large $r/\dt$ the result quickly becomes independent of $\dt$.

\section{Late time behaviour of $\phi^2$}
\label{late2}
In \cite{dgm2}, we found some interesting behaviour for the expectation value of $\phi^2$ when we examined $d=4$ at late times. Essentially, the expectation value for the smooth quench did not depend on the quench duration $\dt$. This  result led us to conjecture that this late time behaviour agrees with that in an instantaneous quench. 

In this section, we return to this issue, first by reviewing the $d=3$ result and then by considering late time behaviour for higher dimensions. We will show that the agreement between smooth and instantaneous quenches found in $d=3$ does not generally occur in higher dimensions.

\subsection{Review of $d=3$}

The starting point will be to consider the correlator in eq.~(\ref{xprime}) for instantaneous quench,  and evaluate this expression at coincident points in space and time, \ie $\vec{x}=\vec{x}'$, $t=t'$. Of course, this expectation value is divergent in the UV, so it must be regulated. In \cite{dgm2} we showed how to carry out the regularization in detail, but for now it will be enough to compute the difference between the quenched expectation value and that for a fixed mass $m$ to produce a finite result. After subtracting, we get the finite difference
\begin{eqnarray}
\langle \phi^2 \rangle_{quench} - \langle \phi^2  \rangle_{fixed} & = & \frac{m^2}{4\pi}  \int \frac{dk}{k \sqrt{k^2+m^2}}\, \sin^2{kt} \label{all_t}
\end{eqnarray}
for $d=3$. %In this expression, we have substituted $\sigma_s=4\pi$ for $d=3$. 
Interestingly, this expression can be integrated analytically and the solution expressed in terms of generalized hypergeometric functions,
\beq
\langle \phi^2 \rangle_{quench} - \langle \phi^2  \rangle_{fixed} = \frac{m^2 t^2}{4\pi} \left(\frac{\pi}{2\,t}  \, _1F_2\left(\frac{1}{2};1,\frac{3}{2};m^2 t^2\right)- \, _2F_3\left(1,1;\frac{3}{2},\frac{3}{2},2; m^2 t^2\right)\right)\,. \label{d3exact}
\eeq

The complete analysis of this solution can be found in \cite{dgm2}, but let us just say here that the expectation value begins by growing linearly when $mt \ll 1$ (but still $t/\dt\gg 1$) but then for very late times, \ie $mt \gg 1$, the expectation value keeps increasing but now only at a logarithmic rate, \ie $\vev{\phi^2}_{ren} \sim \,\log ({m t})$. As we show in \cite{dgm2}, the instantaneous quench and the smooth quench calculation coincide for $d=3$, basically because the integrand for the smooth quench decays fast enough, in a way that the approximation of $\omega \dt \ll 1$ continues to be valid. Recall from the discussion at the end of section \ref{responseq}, that it is possible to obtain the instantaneous quench expectation value starting from the smooth quench and taking both the late time limit, \ie $t/\dt \gg1$, and the low energy limit, \ie $\omega \dt \ll 1$, for every $\omega$ in the problem. However, we generally need to integrate momentum $k$ up to infinity with fixed $\dt$, so usually this approximation will break down for large enough $k$ (remember that $\omega_{out}=k$ in the quench to the critical point). In the special case of $d=3$, though, the integrand decays in a way that only the low momentum modes contribute and then the approximation is reasonable. In the next subsection, we will show that this is a particular effect of the three-dimensional case and does not hold in higher dimensional spacetimes.

\subsection{Higher dimensions}

The main problem that arises in taking the late time limit in higher dimensions is that the expectation value for the instantaneous quench cannot be regulated by simply subtracting the fixed mass expectation value for $d>3$. Moreover we will show that it cannot be regulated using the usual counterterms found in \cite{dgm1, dgm2}. This fact will be taken as a hint to argue that in fact, the low energy approximation is not valid in evaluating the late time expectation value of $\phi^2$ in higher dimensions. Then, in order to get the expectation value for the smooth quench, what one should really do is to fully evaluate eq. (\ref{correlator}) in the limit where $t/\dt \gg 1$. Of course, without taking the extra low energy approximation, it will be impossible to recover the instantaneous quench answer, that it will turn out to be UV divergent for $d\geq 7$ and then, infinitely different from the UV finite result for the smooth quench.

So, let us start by evaluating the bare expectation value for $\phi^2$ in the case of an instantaneous quench. In this case we have,
\beq 
\langle \phi^2 \rangle = \frac{\Omega_{d-2}}{2 (2 \pi)^{d-1}} \int \frac{k^{d-4}\, dk}{\sqrt{k^2 + m^2}} \left( k^2 + m^2\, \sin^2 (kt) \right) \, .
\eeq
To explore the UV behaviour, we expand the expression above for large $k$ and for up to $d=9$, the results can be summarized as 
\begin{eqnarray}
\langle \phi^2 \rangle& =& \frac{1}{\sigma_s} \int dk\, k^{d-4} \left( k -\frac{m^2}{2k} +\frac{3 m^4}{8 k^3} -\frac{5 m^6}{16 k^5} + O(k^{-7})  +  \right. \nnn \\
&& \qquad\left. +  \sin^2 (kt) \left( \frac{m^2}{k} -\frac{m^4}{2 k^3} +\frac{3 m^6}{8 k^5} -\frac{5 m^8}{16 k^7} + O(k^{-9}) \right) \right)\,. \label{div_long_times}
\end{eqnarray}

Of course, the terms appearing in the first line are those same divergent terms that we expect from the constant mass case,
\beq
\langle \phi^2 \rangle_{fixed} = \frac{1}{\sigma_s} \int \frac{k^{d-2}}{\sqrt{k^2 + m^2}} dk\, .
\label{phi_sq_fixed}
\eeq

But in eq. (\ref{div_long_times}), we also have divergent terms in the  second line proportional to $\sin^2(kt)$ that do not correspond to any physical counterterm contributions found in eq.~\reef{ict}. In fact, as we are interested in the long time behaviour of the expectation value and given that the mass profile (and its time derivatives) decay exponentially in time, the only remaining physical counterterm in this limit should be the mass independent term of eq. \reef{ict}, \ie $k^{d-3}$.

\begin{figure}[H]
        \centering
        \subfigure[$\Phi^2(k)$ for small momenta.]{
                \includegraphics[scale=0.55]{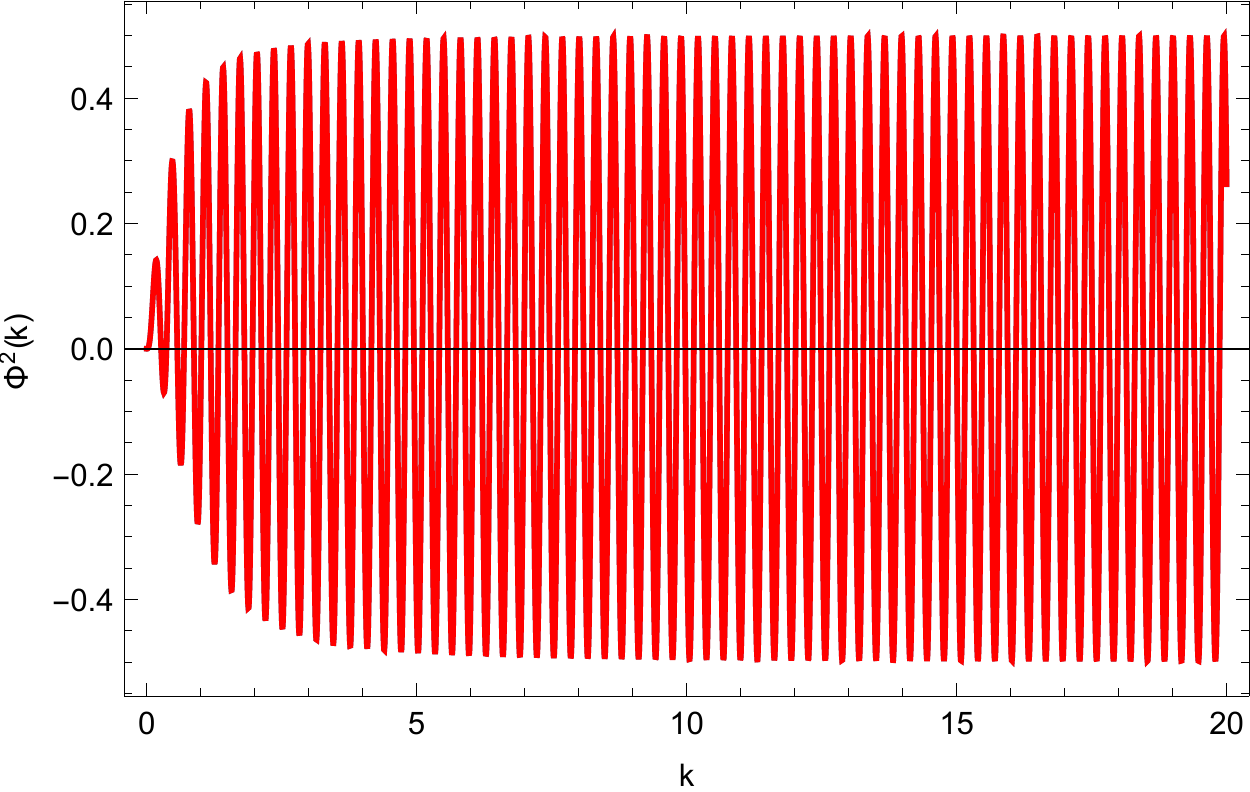} \label{d5_approx_a}}
   		 \subfigure[$\Phi^2(k)$ for large momenta.]{
                \includegraphics[scale=0.55]{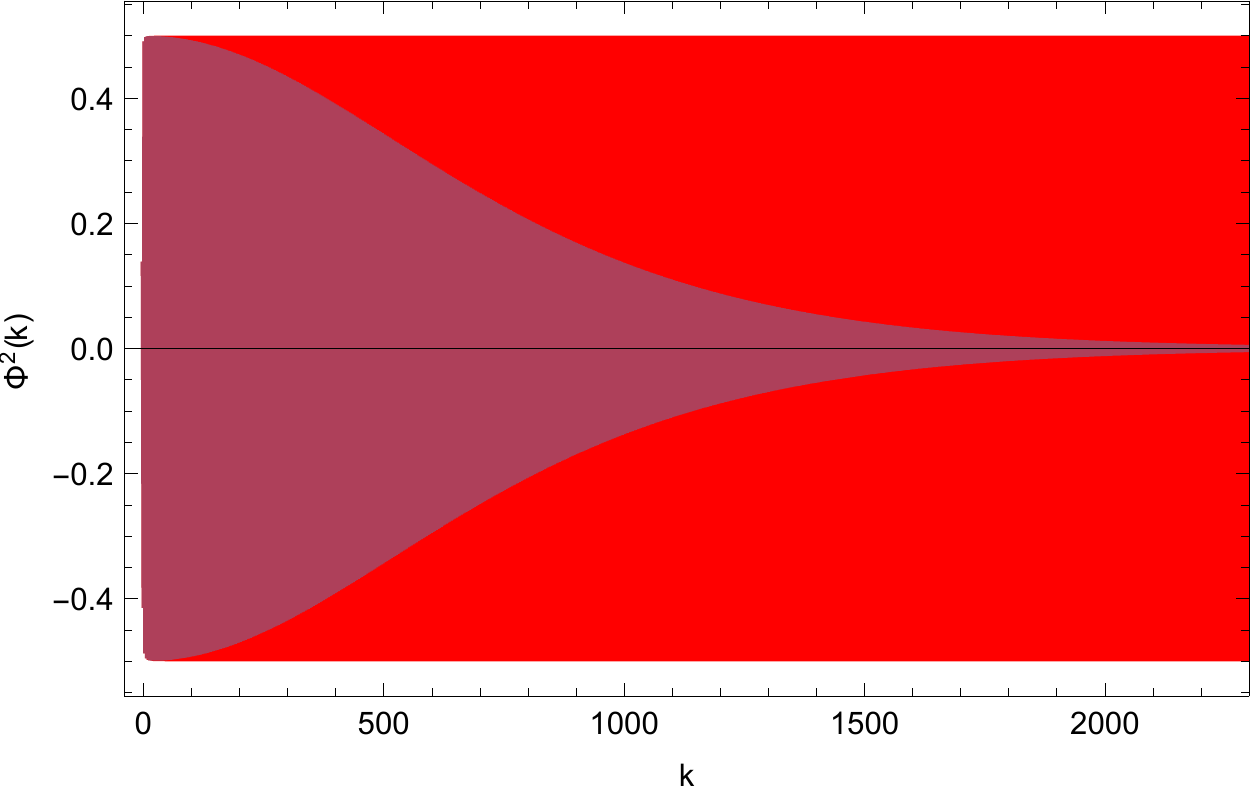} \label{d5_approx_b}}
        \caption{(Colour online) Analysis of the approximation of low energies and late times in $d=5$. We show results for $\dt=10^{-3}$, $t=10$ and $m=1$. In the first figure, we show that $\Phi^2(k)$ in both eqs. (\ref{full_late}) and (\ref{longtime_full2}) coincide when evaluated for small $k$. However in the second plot, we see that the full integrand (blue) decays for large $k$ while the approximate solution (red) keeps oscillating. The second plot looks fully painted because of the highly oscillatory nature of the functions.} \label{fig_approx_d5}
\end{figure}

\begin{figure}[H]
        \centering
        \subfigure[$\Phi^2(k)$ for small momenta.]{
                \includegraphics[scale=0.55]{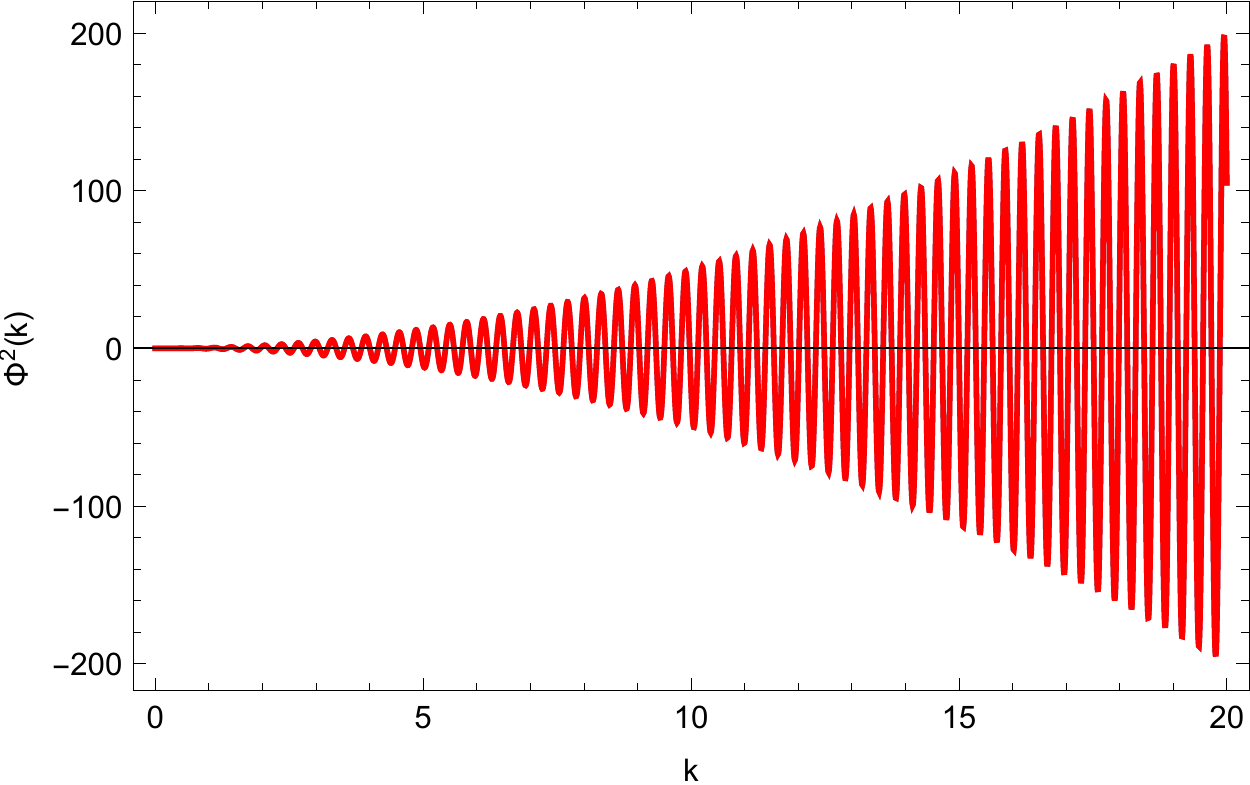} \label{d7_approx_a}}
   		 \subfigure[$\Phi^2(k)$ for large momenta.]{
                \includegraphics[scale=0.55]{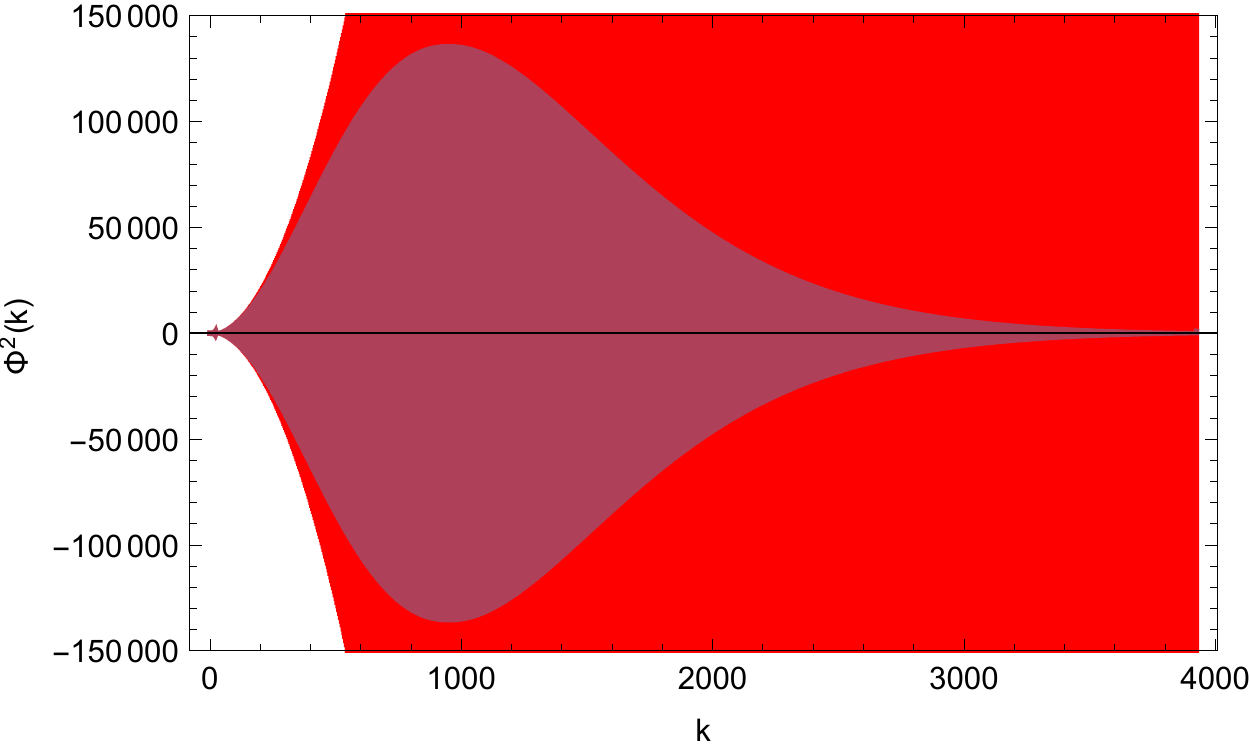} \label{d7_approx_b}}
        \caption{(Colour online) Analysis of the approximation of low energies and late times in $d=7$. We show results for $\dt=10^{-3}$, $t=10$ and $m=1$. In the first figure, we show that $\Phi^2(k)$ of both eqs. (\ref{full_late}) and (\ref{longtime_full2}) coincide when evaluated for small $k$. However in the second plot, we see that the full integrand (blue) decays for large $k$ while the approximate solution (red) keeps oscillating. The second plot looks fully painted because of the highly oscillatory nature of the functions.} \label{fig_approx_d7}
\end{figure}

In evaluating $\langle \phi^2 \rangle_{ren}$, we integrate over all momenta, but the integral is divergent (even after taking into account the physical counterterms). The reason for this behaviour is that in higher dimensions the approximations of eq. (\ref{2-1}) are not longer valid. One way to realize this is to compare it with the expression before that approximation. Recall that this is given by eq. (\ref{correlator}), which after some manipulation, in the late time limit becomes,
\bea
\langle \phi^2 \rangle_{smooth} & = & \sigma_s^{-1} \int \Phi^2(k) \, dk \label{full_late}\\
& = & \sigma_s^{-1} \int dk \left(\frac{k^{d-2}}{\omega_{out}} \Big\{ |\alpha_\vk|^2 + \alpha_\vk \beta^\star_{-\vk}~e^{2 i \w_{out} t}+
\alpha^\star_\vk \beta_\vk ~e^{-2 i \w_{out} t} +|\beta_\vk|^2 \Big\} - k^{d-3} \right), \nonumber
\eea
where $\alpha_\vk$ and $\beta_\vk$ are given by eq. (\ref{bogocoeff}). Now we wish to compare this integral with the result for an instantaneous quench
\beq
\langle \phi^2  \rangle_{instant} = \sigma_s^{-1} \int \Phi^2(k) \, dk = \sigma_s^{-1} \int dk \left( \frac{k^{d-4}}{\sqrt{k^2 + m^2}} \left( k^2 + m^2 \sin^2(kt) \right) - k^{d-3} \right) \, . \label{longtime_full2}
\eeq
For this comparison, we start by examining the integrands $\Phi^2(k)$. For $d=5$, this is done in fig.~\ref{fig_approx_d5}, where we choose $\dt=10^{-3}$ and we evaluate the expression at $t=10$. However, our results generalize to the full range of values where the approximation of late times is valid. What we see is interesting: if we focus on small momenta, as shown in fig.~\ref{d5_approx_a}, both the approximate and the full integrands agree. They both show a highly oscillating behaviour, that seems to continue to larger momenta and which would make both integrals  diverge if it did so. However, what we see in fig.~\ref{d5_approx_b} is that in fact this behaviour does not continue for very large $k$ in the case of eq. (\ref{full_late}). It can be seen that $\Phi^2(k)$ decays to zero for large momentum in the case of the full integral, as required to produce a UV finite result. Instead, the approximate integrand continues to oscillate and so produces a UV divergent integral. In fig. (\ref{fig_approx_d5}), we see that the two integrands differ substantially for $k\gtrsim1000$, \ie for $k \dt \gtrsim 1$. Of course, the approximation of $\w_{out} \dt \ll 1$ is no longer valid in this regime and hence it is natural to expect that they should differ there. The approximations of eq. (\ref{2-1}) are not valid to obtain the correct late time limit of $\langle \phi^2 \rangle$ in higher dimensions and hence the expectation value does not match that after an instantaneous quench. To complete this analysis we show that the same happens for $d=7$ in fig.~\ref{fig_approx_d7}, where the oscillatory behaviour is even increased by a power law divergence in the approximate integrand.

To conclude this section, let us summarize our findings with regards to the late time limit after the quench. We have considered two different protocols to quench a scalar field: the instantaneous one, where we suddenly start evolving an eigenstate of the massive case with the massless Hamiltonian; and the smooth one, where we continuously evolve the mass of the scalar field from the massive case to the massless in a time scale of $\dt$. Now one would think that these different protocols must give different answers in the early time evolution,\footnote{In our case, we obtain this interesting universal scaling near $t=0$.} but that in the limit of $\dt$ going to zero and for late times, we should obtain similar results. In \cite{dgm2} we showed that effectively, if we take the limit of $t/\dt \gg 1$ \textit{and also} $\omega\dt \ll1$ for every $\omega$, we can reproduce the instantaneous quench result from that for the smooth quench and so in principle, we might expect the same late time behaviour in both cases. This allowed us to identify the interesting logarithmic growth behaviour of the scalar field for late times in $d=3$. In a self consistent way, we showed that both of these approximations were reasonable in $d=3$ and so, the late time behaviour for the smooth and the instantaneous quench agreed. However, we found that this agreement in $d=3$ was fortuitous because when we tried to repeat the analysis in higher dimensions, we found that the approximations of eq. (\ref{2-1}) are no longer valid. That is, higher momenta (and hence, higher frequencies) contribute significantly to the expectation value of $\phi^2$ in higher dimensions and cannot be neglected. Hence for $d\geq 4$, the instantaneous quench gives a result for $\langle \phi^2 \rangle$ at late times which is infinitely different from the smooth quench result. In particular, the smooth quench gives a finite late time limit for $\vev{\phi^2}$ as $\dt \to 0$ \cite{dgm2}, while the corresponding result appears to diverge in an instantaneous quench.

\subsection{Regulated instantaneous quench}

It is interesting to note that the integrand in eq. (\ref{longtime_full2}) for the instantaneous quench in higher dimensions does not decay to zero for large momentum. Instead, it \textit{seems} to show a rapid oscillatory behaviour around zero, as shown in figs. (\ref{fig_approx_d5}) and (\ref{fig_approx_d7}). So one may think that even though the amplitude is diverging, the positive and negative part are cancelling in every period and so, in some sense, we may be able to recover a finite result from these integrals. In fact, we are inspired here by the way that the fixed mass  correlators were regulated in Appendix \ref{const_mass_corr}.

\begin{figure}[h!]
\setlength{\abovecaptionskip}{0 pt}
\centering
\includegraphics[scale=0.8]{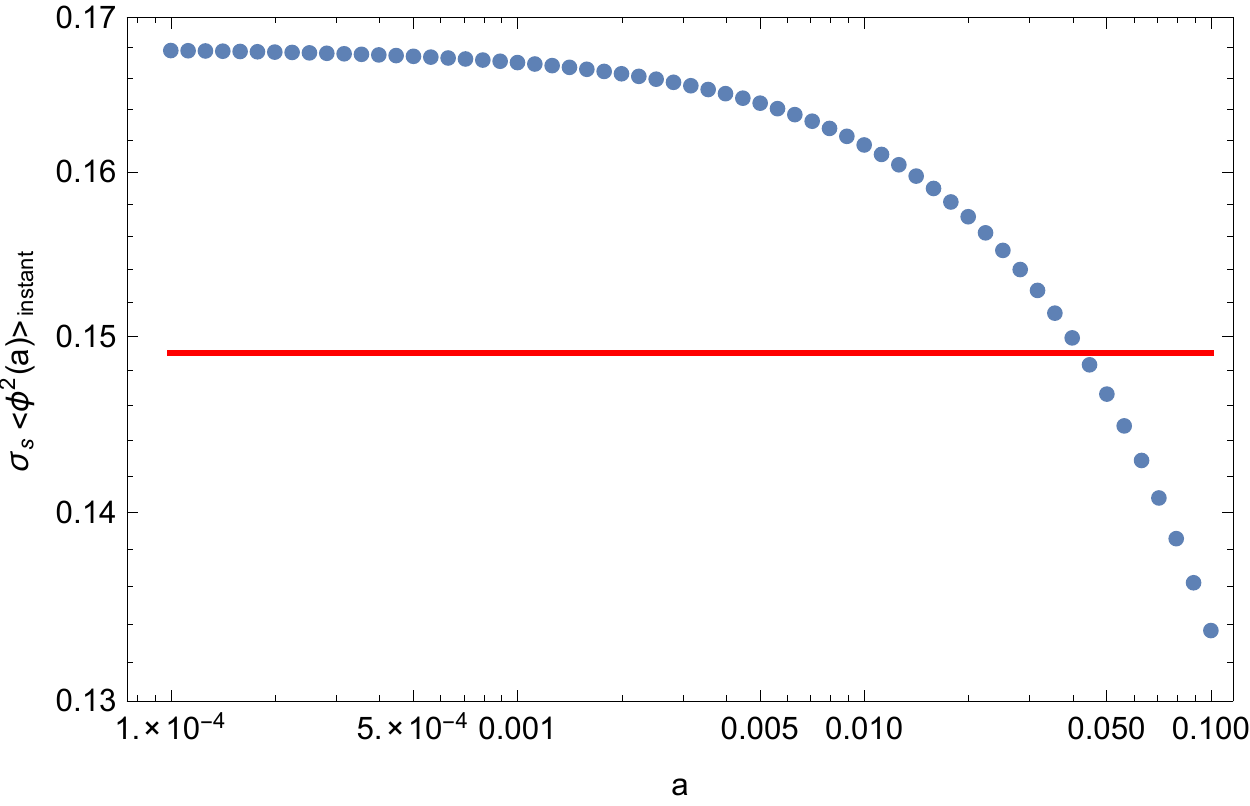}
\caption{(Colour online) Expectation value of $\phi^2$ in the instantaneous quench as a function of the regulator parameter $a$ for $d=5$ and $m t=10$. The values of $a$ go from $10^{-1}$ to $10^{-4}$ in a logarithmic scale. The solid line represents the value for the smooth quench at $mt=10$ and $\dt=1/20$.} \label{fig_d5_extra_reg}
\end{figure}

Hence let us go back to our instantaneous quench results. For $d=5$, the behaviour of the expectation value of $\phi^2$ for large $k$ can be extracted from eq. (\ref{div_long_times}). This gives,
\beq
\langle \phi^2 \rangle \xrightarrow[k\to \infty]{} \frac{1}{\sigma_s} \int dk\, \left( k^2 -\frac{m^2}{2} + m^2 \sin^2 (k t) \right)\,. \label{div_long_times_d5}
\eeq
The first divergence, proportional to $k^{d-3}$ is our usual counterterm that we will subtract. But note, then, that the term proportional to $\sin^2$ can be re-written to yield
\begin{eqnarray}
\langle \phi^2 \rangle_{ren} \xrightarrow[k\to \infty]{} \frac{1}{\sigma_s} \int dk\, \left( -\frac{m^2}{2} + m^2 \frac{1- \cos(2 k t)}{2} \right) = -\frac{1}{\sigma_s} \int dk\, \frac{\cos(2 k t)}{2}\, , \label{div_long_times_d5_v2}
\end{eqnarray}
and this is one of the integrals that we now know how to regulate (just put $\alpha =0$ and $x=2t$ in eq. (\ref{trig_reg}). Of course, this is just showing the large $k$ behaviour of the integral. In order to get the full answer we should include all momenta and this, unfortunately, can only be done numerically. But, in principle, since we know how the integral behaves for large $k$, we should be able to get a finite result for our integral as well by using this new regulator. The results are shown in fig.~\ref{fig_d5_extra_reg}, where we evaluate numerically the instantaneous quench solution of eq. (\ref{longtime_full2}) for $d=5$ and adding a regulator $\exp(-a k)$,
\beq
\langle \phi^2 (a) \rangle_{instant} = \sigma_s^{-1} \int dk \left( \frac{k}{\sqrt{k^2 + m^2}} \left( k^2 + m^2 \sin^2(kt) \right) - k^{2} \right) \exp(-a k) \, .
\eeq

We evaluate the expectation value at late times, $m t=10$, and compare it with the smooth quench also at late times, where we are using $\dt = 1/20$. We find that as we take $a$ to zero, the regulated integral for the instantaneous quench approaches some finite value, showing that the integral converges. However, this value differs from that for the smooth quench in a relative amount by
\beq
\left| \frac{\langle \phi^2 (a\to0) \rangle _{instant} - \langle \phi^2 \rangle _{smooth}}{\langle \phi^2 \rangle _{smooth}}\right| \sim \left| \frac{0.1670-0.1490}{0.1490} \right| \sim 0.121.
\label{difference}
\eeq
So even if we found a way to make sense of the divergent integral in $d=5$, the result does not quite coincide with the smooth quench result, where no approximation is made (apart from the late time limit approximation).

\begin{figure}[h!]
\setlength{\abovecaptionskip}{0 pt}
\centering
\includegraphics[scale=0.8]{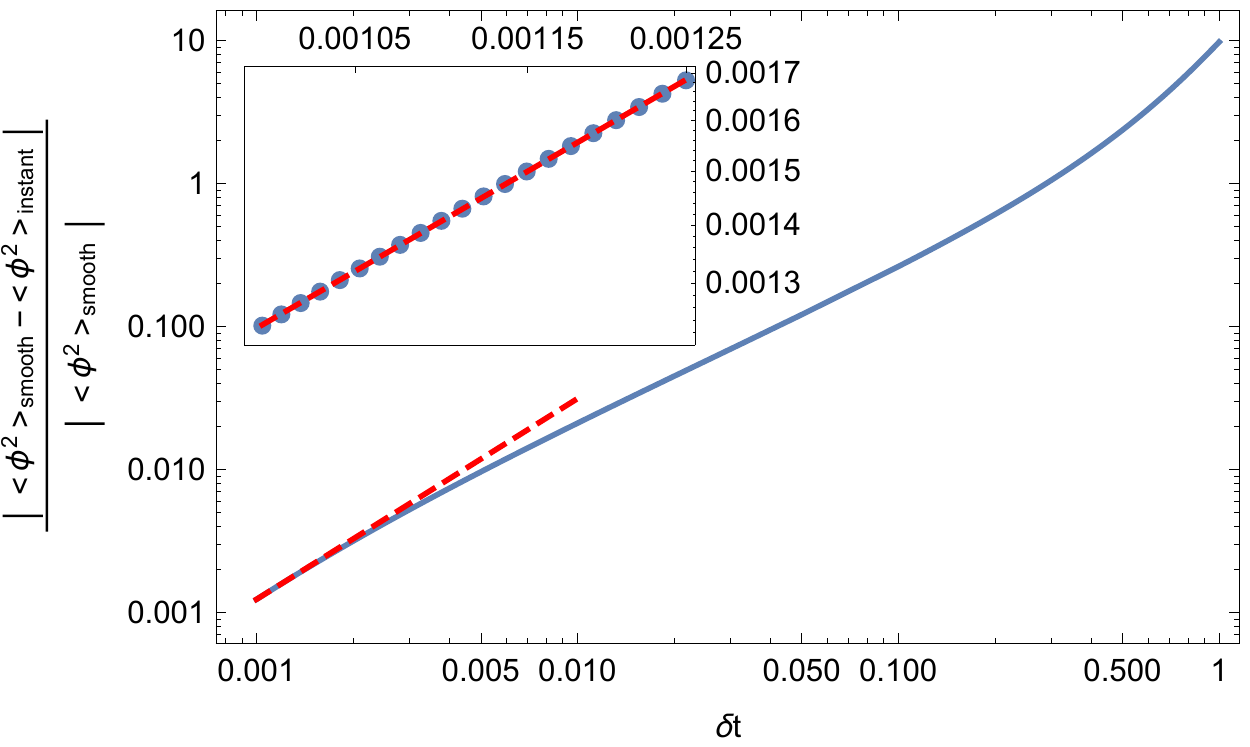}
\caption{(Colour online) Relative error in the computation of $\vev{\phi^2}$ at late times using the smooth and the instantaneous quench as a funtion of $\dt$. Here we are regulating the instantaneous result with $a=10^{-3}$ and evaluating at time $mt=10$. The red dashed line is showing the fit of the first points (from $\dt=1/1000$ to $\dt=1/800$) by a power law function of the type $f(\dt)=a \dt^\alpha$, where $a=20.03$ and $\alpha = 1.404$. The inset zooms in the region where the fit was made showing perfect agreement between the points and the fit.}\label{rel_error}
\end{figure}

Note that the relative difference in eq. (\ref{difference}) is of order $m \dt$. To quantify this difference more precisely we compute the expectation value in the instantaneous quench case with $a=10^{-3}$ and $mt=10$. We then vary $\dt$ in the case of the smooth quench and compute the relative error between the two results. The outcome is shown in fig.~\ref{rel_error}. We see that for $\dt \sim 1$, the two protocols give very different answers. But this is expected because a large $\dt$ means going into the adiabatic regime and this need not to agree with the rapid quench even at late times. Instead, as we decrease $\dt$, we see that the relative difference between the two approaches also diminishes and in fact, when $\dt$ is of order $10^{-3}$, the relative error is also of that order of magnitude.

Naively, one may think that it is possible to understand this behaviour by expanding the expectation value in powers of $\dt$. To do that we start with the smooth quench integral in eq. (\ref{full_late}). We then expand the Bogoliubov coefficients $\alpha_\vk$ and $\beta_\vk$ for small $\dt$ and compute the integrand to lowest orders in $\dt$. This results in
\bea
\vev{\phi^2}_{smooth}=\frac{1}{\sigma_s} \int & dk & \frac{k^{d-4}}{\sqrt{k^2+m^2}} \left(  \left( k^2 + m^2 \sin(kt)^2 \right) + \right. \label{smoothseries}\\
& & \left. + \dt^2 \frac{\pi^2 m^2}{12} \left( k^2 - (2 k^2 + m^2) \sin ^2(k t) \right) + O(\dt^3) \right) \,.\nonumber
\eea
Let us analyse this last expression. The first term is independent of $\dt$ and one can easily see that it exactly matches the expression for the instantaneous quench in eq. (\ref{longtime_full2}). Of course, this match was known implicitly, since expanding for $\dt\ll 1$ in the previous expression is actually expanding for $\omega \dt \ll 1$ and the agreement found above is the claim that we could reproduce the instantaneous result by taking the small frequency limit of the smooth quench. The next term in the $\dt$ expansion appears at order $\dt^2$. So, again naively, one might expect that $\vev{\phi^2}_{smooth} = \vev{\phi^2}_{instant} + \gamma \dt^2 + O(\dt^3)$, for some number $\gamma$. However, if we look carefully at the integral that gives this correction at order $\dt^2$, we will see that it is in fact divergent for $d\geq 5$. We can try to regulate it by adding a regulator as in Appendix \ref{const_mass_corr}, but in fact we will see that apart from the oscillating term (which can be regulated) there is an extra constant term in the integrand, \ie $-\frac{1}{24 \sigma_s} \pi ^2 \dt^2 m^4$, that will make the integral divergent as we integrate over $k$ from 0 to $\infty$.

This is closely related to the fact that if we try to fit the relative error by some power law expression in the region of small $\dt$ --- see red dashed line in fig.~\ref{rel_error} --- we find that the error does not scale as $\dt^2$ but the exponent is rather close to $1.40$. This is another sign that this naive expansion is somehow ill-defined for $d=5$. Again, we should say that behind this expansion it is assumed not only that $\dt$ is small but that $\omega \dt \ll 1$, for every $\omega$ in the problem, and we already showed that the assumption is not valid for $d\geq 5$.

The situation is even worse in higher dimensions, where the first term in the series, \ie the order $\dt^0$ term in eq. (\ref{smoothseries}), fails to converge. As an example, we show what happens in $d=7$. Even though in fig.~\ref{fig_approx_d7} the integrand appears to oscillate around zero, this is not the case. For large $k$, we have, after subtracting the usual counterterms,
\begin{eqnarray}
\langle \phi^2 \rangle_{ren} \xrightarrow[k\to \infty]{} & & \frac{1}{\sigma_s}  \int dk \, \left( -\frac{m^2}{2} k^2 + \frac{3 m^4}{8} + \left( m^2 k^2 - \frac{m^4}{2} \right) \sin^2(kt) \right)\nnn  \\
& & \simeq \frac{1}{\sigma_s} \int dk \, \left( \frac{1}{4} \left(m^4-2 m^2 k^2\right) \cos (2 k t)+\frac{m^4}{8}\right).  \label{div_long_times_d7_v2}
\end{eqnarray}
So, even though the first two terms could be regulated using the above prescription, there is an extra constant term that cannot. Adding the regulator to the term proportional to $m^4/8$, we will get a result which diverges as $a\to0$,
\beq
\frac{m^4}{8} \int_0^{\infty} \exp(-a k) = \frac{m^4}{8 a},
\eeq
and so, the limit of $a\to0$ in this case will be nonsense. The same also happens in any higher number of dimensions. In fact, the only reason why this worked in $d=5$ was because the divergent  terms with the $\sin^2$ term, exactly matched the terms without the $\sin^2$ term, in a way that made the whole integrand to oscillate around zero. But this, as shown above, does not happen in general and so, the instantaneous quench approximation in higher dimensions gives a value for $\langle \phi^2 \rangle$ which differs by an infinite amount from the smooth quench result, even in the late time limit.

\section{The energy density at late times}
\label{energylate}

In \cite{dgm1,dgm2}, it was argued that the renormalized energy density for a quench satisfying eq. (\ref{0-2}) obeys a scaling relation (\ref{0-3}),
\ben
\delta \cale_{ren} \sim \delta \lambda^2 (\delta t)^{d-2\Delta}\,.
\label{0-3a}
\een
This result is consistent with the scaling of the expectation value of the operator (\ref{0-4}) since it satisfies the Ward identity
\ben
\frac{d\cale_{ren}}{dt} = - \, \partial_t \lambda (t) \, \langle \calo_\Delta \rangle_{ren}\,.
\label{5-1}
\een
In the corresponding scaling relation (\ref{0-4}) for the quenched operator, 
$ \langle \calo_\Delta \rangle_{ren}$ is measured earlier than or soon after the end of the quench. However, the energy scaling (\ref{0-3a}) will be valid for arbitrarily late times since the energy is injected into the system only during the quench. 
The equation (\ref{0-3a}) gives the $\dt$ dependence. The energy density itself will have additional finite pieces, which would be subdominant for $\Delta > d/2$, but in fact give the dominant contribution for $\Delta < d/2$.

In this section, we will concentrate on the energy density for the free bosonic field with the mass profile (\ref{massprofile}) at asymptotically late times and calculate a UV finite quantity: the excess energy above the ground state energy of the system with the value of the coupling at asymptotically late times. We will perform the $\dt \rightarrow 0$ limit and compare the results with that of an instantaneous quench. 

In terms of the ``in" modes the energy density is given by
\ben
\ce = \frac{1}{2}\int \frac{d^{d-1}k}{(2\pi)^{d-1}}\left( |\partial_t u_\vk|^2 + (k^2 + m^2(t))|u_\vk|^2 \right)
\label{5-2}
\een
Since we are interested in the behavior of this quantity at late times, it is convenient to express this in terms of the ``out" modes. In terms of the Bogoliubov coefficients $\alpha_\vk, \beta_\vk$ defined in (\ref{bogo}) and (\ref{bogocoeff}) this becomes 
\bea
\ce = \frac{1}{2}\int \frac{d^{d-1}k}{(2\pi)^{d-1}}& & ( [\omega_{out}^2 + k^2 + m^2(t)](|\alpha_\vk|^2 + |\beta_\vk|^2) |v_\vk|^2| \nnn \\
& & +  [k^2 +m^2(t)-\omega_{out}^2](\alpha_\vk \beta^\star_\vk v_\vk v_{-\vk} + \alpha^\star_\vk \beta_\vk v^\star_\vk v^\star_{-\vk}) ) \,.
\label{5-3}
\eea
In deriving this expression, we have used the fact that $\alpha_\vk$ and $\beta_\vk$ depend only on $|\vk|$. At $t = \infty$, the second line of eq. (\ref{5-3}) vanishes. Using the relation $|\alpha_\vk|^2 - |\beta_\vk|^2 =1$ and the asymptotic form of the out modes at $t \rightarrow \infty$, \ie
\ben
v_\vk  \rightarrow  \frac{1}{\sqrt{2\omega_{out}}}e^{i(\vk \cdot \vx - \omega_{out}t)}\,,
\label{5-4}
\een
one gets
\ben
\ce = \frac{1}{2}\int \frac{d^{d-1}k}{(2\pi)^{d-1}} \omega_{out} \left( 1 + 2 |\beta_\vk|^2 \right)\,.
\label{5-5}
\een
However, the ground state energy of the system with the final value of the mass is given by
\ben
\ce_{ground} = \frac{1}{2}\int \frac{d^{d-1}k}{(2\pi)^{d-1}}\omega_{out} \,.
\een
Therefore the excess energy over the final ground state is given by
\ben
\Delta \ce = \int \frac{d^{d-1}k}{(2\pi)^{d-1}}\omega_{out} |\beta_\vk|^2 \,.
\label{5-6}
\een
Using the explicit form of the Bogoliubov coefficients in eq. (\ref{bogocoeff}) and integrating over the angles, we arrive at the final expression
\ben
\Delta \ce = \frac{\Omega_{d-2}}{(2\pi)^{d-1}} \int_0^\infty dk~k^{d-2}~ \omega_{out} \frac{\sinh^2 (\pi \omega_- \dt)}{\sinh (\pi \omega_{in} \dt) \sinh (\pi \omega_{out} \dt)} \,.
\label{5-7}
\een
For the mass profile (\ref{massprofile}), the integral in eq. (\ref{5-7}) is finite for any finite $\dt$. In fact, for small $k$, the integrand approaches $(k^{d-2}) \frac{\tanh(\pi m \dt)}{2\pi\dt}$, while for large $k$, it becomes $(k^{d-3})m^4 (\dt)^2 e^{-2\pi k \dt}$. Hence the integral above is convergent both in the IR and UV for any physical $d \geq 2$. 

\begin{figure}[h!]
\setlength{\abovecaptionskip}{0 pt}
\centering
\includegraphics[scale=0.8]{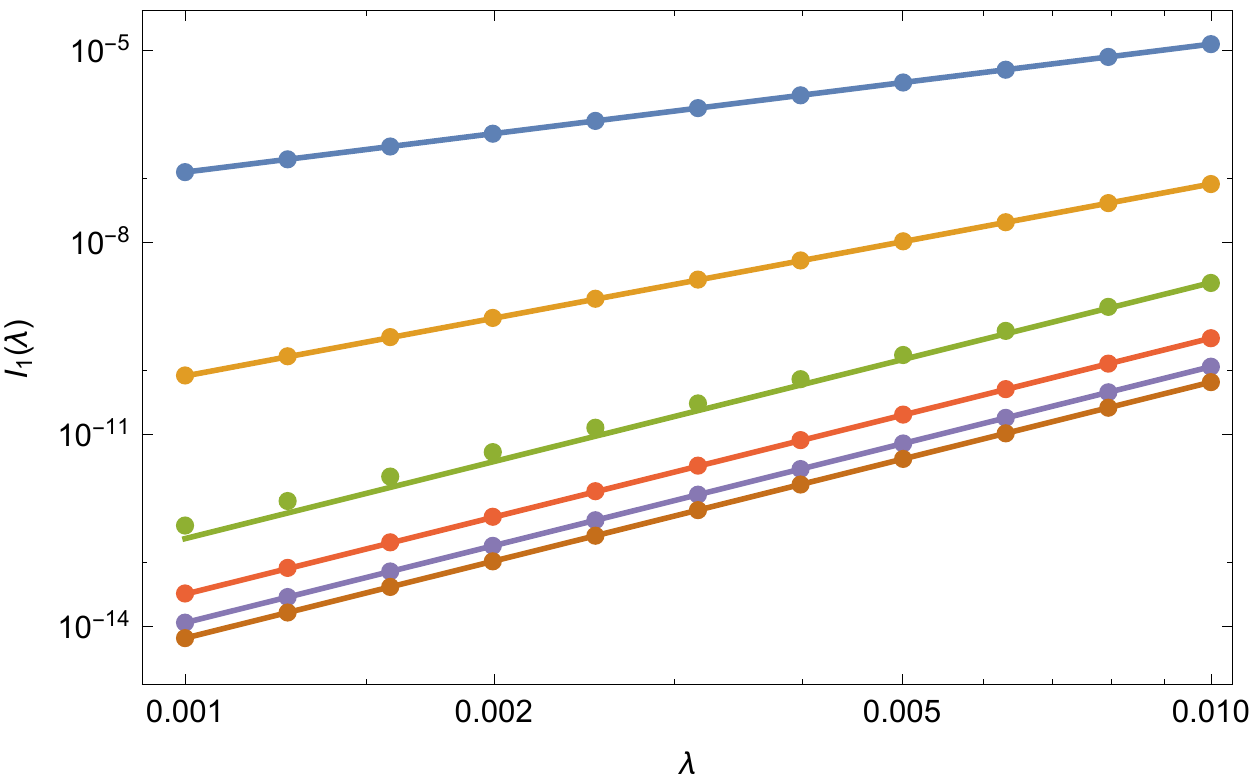}
\caption{(Colour online) The integral $I_1(\lambda)$ in eq. (\ref{5-7-1}) as a function of $\lambda$ on a logarithmic scale. The data are for $d=2,3,4,5,6,7$ from top to bottom. The solid lines correspond to $I_1(\lambda) \sim \lambda^d$ for $d=2,3$ and $I_1(\lambda) \sim \lambda^4$ for $d=4,5,6,7$. Note that the fit in $d=4$ is not as good as the others, which suggests that there are log corrections to the leading behavior for $d=4$.} \label{fig_I1}
\end{figure}

To analyze the small $\lambda = m\dt$ limit of eq. (\ref{5-7}), let us first scale out a power of $\dt$ to write $\Delta \ce = (\dt)^{-d} I_1(\lambda)$ where
\ben
I_1(\lambda) = \int_0^\infty dq~q^{d-1}~\frac{\sinh^2 [\frac{\pi}{2}(\sqrt{q^2+\lambda^2}-q)]}{\sinh(\pi\sqrt{q^2+\lambda^2})\sinh(\pi q)} \,.
\label{5-7-1}
\een
Clearly $I_1(0)=0$. However the small $\lambda$ dependence is different for different dimensions. It turns out that
\bea
I_1(\lambda) & \sim & \lambda^{d}~~~~d=2,3 \,, \nonumber \\
& \sim & \lambda^4~~~~d \geq 4 \,.
\label{5-7-2}
\eea
The above behavior was determined from a direct numerical evaluation of the integral in eq.  (\ref{5-7-1}) and fitting the results shown in fig.~\ref{fig_I1}.

\begin{figure}[h!]
\setlength{\abovecaptionskip}{0 pt}
\centering
\includegraphics[scale=0.8]{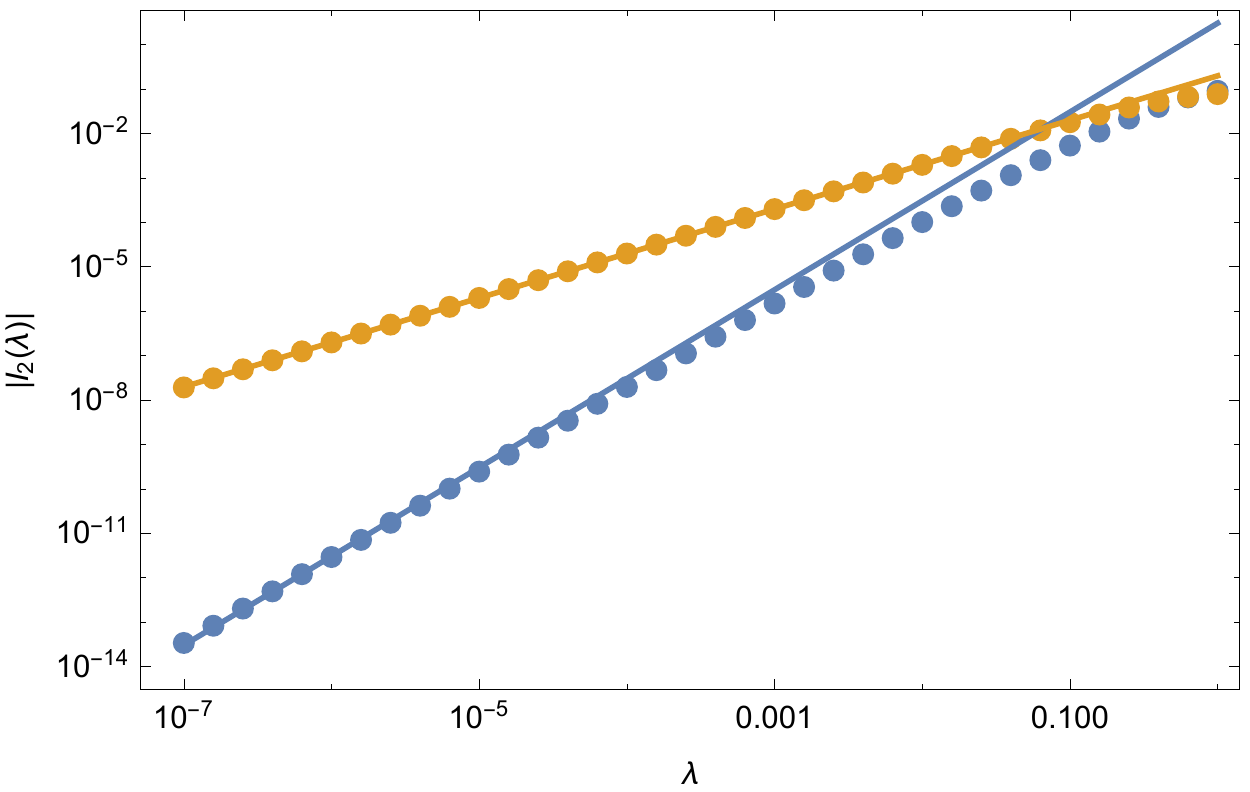}
\caption{(Colour online) The absolute value of $I_2(\lambda) = m^{-d} \, \delta\ce$ as a function of $\lambda=m\,\dt$ on a logarithmic scale. The data are for $d=2,3$ from bottom to top. The solid lines correspond to the curves $I_2(\lambda) \sim \lambda^2$ for $d=2$ and $\lambda$ for $d=3$. Those curves correspond to the best fit of the numerical data in the region $10^{-7}<\lambda<10^{-5}$.} \label{deltaI}
\end{figure}

This means that for $d=2,3$, the excess energy has a smooth finite small $\dt$ limit with $\Delta \ce \sim m^d$. The leading answer is exactly the same as the energy excess for the instantaneous quench, which can be read off easily from the corresponding Bogoliubov coefficients (\ref{bogocoeff2}) (for the mass profile (\ref{massprofile}))
\ben
\Delta \ce^{instant} =  \frac{\Omega_{d-2}}{(2\pi)^{d-1}} \int dk~k^{d-1}
\frac{(\sqrt{k^2+m^2}-k)^2}{4k\sqrt{k^2+m^2}} \,.
\label{5-9}
\een
This quantity $\Delta \ce^{instant}$ is finite for $d=2,3$ and hence we have
\bea
\Delta \ce^{\dt \rightarrow 0}|_{d=2} & = & \frac{m^2}{16\pi} \,, \nnn \\
\Delta \ce^{\dt \rightarrow 0}|_{d=3} & = & \frac{m^3}{24\pi} \,.
\label{5-10}
\eea
To estimate the corrections to this leading small $\lambda$ behavior consider the difference of the excess energy to the excess energy due to an instantaneous quench,
\ben
\delta \ce = \Delta \ce - \Delta \ce^{instant}= \dt^{-d} \left( I_1(\lambda) - \dt^d \Delta\ce^{instant}\right) \equiv m^d \, I_2(\lambda) \,.
\label{5-10-1}
\een
A numerical evaluation of this quantity shows
\ben
\delta \ce \sim m^4 \, \dt^{4-d} \,,
\label{5-10-2}
\een
both for $d=2$ and $3$. This behaviour is shown in fig.~\ref{deltaI}, which is a log-log plot of the quantity $I_2(\lambda)$. Clearly, we have $I_2(\lambda) \sim \lambda^2$ for $d=2$ and $I_2(\lambda) \sim \lambda$ for $d=3$, which leads to the scaling in eq. (\ref{5-10-2}).

On the other hand, for $d =4,5,\ldots$, one recovers the desired scaling from the leading behaviour, \ie
\ben
\Delta \ce \sim \lambda^4 \, \dt^{-d} = m^4 \, \dt^{4-d} \,,
\label{5-10-3}
\een
with logarithmic corrections for even dimensions.
Thus for these dimensions, the energy density diverges in the $m\dt \rightarrow 0 $ limit. At the same time, the energy density for an instantaneous quench diverges in the UV for these dimensions. Let us emphasize this point once again: our results show that the energy density is UV finite after a smooth quench; however, in the limit $m \dt \to 0$, that energy density diverges, just as in the instantaneous quench, where the divergence is in the UV.

These results also indicate that for $d \geq 4$, the excess energy is in fact given by linear response. This is in accord with the computation of the renormalized expectation value of $\phi^2$ and the renormalized energy density for $d \geq 4$, as described in \cite{dgm1, dgm2}. Indeed if we extract the leading piece in a small $\lambda$ expansion by expanding the integrand in eq. (\ref{5-7-1}) we get the expression 
\ben
I_1(\lambda) \sim \frac{\pi^2 \lambda^4}{16}\int_0^\infty dq~\frac{q^{d-3}}{(\sinh(\pi q))^2} \,.
\een
This expansion makes sense when the above integral is convergent. However the integral is IR divergent for $d=2,3,4$. The latter explains is why $I_1(\lambda)$ does not have an expansion in terms of $\lambda^2$ in these dimensions.

\section{Excess Energy for General Theories}
\label{general}

The discussion in section \ref{energylate} suggests that the scaling of the excess energy should be a property of a general interacting field theory. In this section, we argue that this is indeed true.

Consider an action
\ben
S = S_{CFT} + \int dt \, \lambda (t) \int d^{d-1} x~\calo (\vx,t) \,,
\label{7-1}
\een
where $S_{CFT}$ is a conformal field theory action.
The function $\lambda (t)$ is of the form
\[ \lambda(t) = \left \{
\begin{array}{ll}
\lambda_0 &\ \ \ {\rm for}\ t < 0\,,\\
\lambda_0 + \delta \lambda\, F(t/\delta t) &\ \ \  {\rm for}\ 0\le t\le \dt\,,\\ 
\lambda_1=\lambda_0 + \delta \lambda\, &\ \ \  {\rm for}\  t> \dt\,. 
\end{array} \right.
\]
Alternatively, we may write $\lambda(t) = \lambda_0 + \delta \lambda \, F(t/\dt)$ if we specify $F(y\le0)=0$ and $F(y\ge1)=1$. We leave the details of the function $F(y)$ during the transition (\ie $0\le y\le 1$) unspecified other than that the maximum is finite with $F_{max}\ge1$.
Further, this profile may dip below zero by some finite amount and so we specify the minimum as $F_{min}\le0$.
Implicitly, we are also assuming that the profile is smooth.  

The system is prepared in the ground state of the initial action. Let us evaluate the total energy density $\ce (t)$ at some time $t > \dt$ in a perturbation expansion in $\delta \lambda$. To quadratic order in $\delta \lambda$, this expression is given by
\bea
\ce (t) & = & \ce_0  - \delta \lambda \, F(t/\dt) \, \langle 0,\lzero| \calo (\vec{0},0) |0,\lzero \rangle   \nonumber \\
& & - \, i \, \delta \lambda^2 \, F(t/\dt) \int_0^t dt^\prime \int d^{d-1}x^\prime  \, F(t^\prime/\dt) \,  \langle 0,\lzero|[ \calo(\vx^\prime,t), \calo (\vec{0},t^\prime)] |0,\lzero \rangle \nonumber \\
& & - \, \ce_0 \, \delta\lambda^2 \, \int_0^t dt^\prime \int_0^{t^\prime} dt^{\prime\prime} \int d^{d-1}x^\prime \, F (t^\prime/\dt) \, F (t^{\prime\prime}/\dt) \, \langle 0,\lzero| \calo(\vx^\prime,t^\prime)\calo(\vec{0},t^{\prime\prime}) |0,\lzero \rangle \nonumber \\
& & - \, \ce_0 \, \delta\lambda^2 \, \int_0^t dt^\prime \int_0^{t^\prime} dt^{\prime\prime} \int d^{d-1}x^\prime \, F (t^\prime/\dt) \, F(t^{\prime\prime}/\dt) \,
\langle 0,\lzero|  \calo(\vec{0},t^{\prime\prime})\calo(\vx^\prime,t^\prime) |0,\lzero \rangle \nonumber \\
& & + \, \delta\lambda^2 \, \int_0^t dt^\prime \int_0^{t} dt^{\prime\prime} \int d^{d-1}x^\prime F(t^\prime/\dt) \, F(t^{\prime\prime}/\dt) \,  \langle 0,\lzero|\calo(\vx^\prime,t^\prime) H_0 \, \calo(\vec{0},t^{\prime\prime}) |0,\lzero\rangle 
\nonumber \\
& & + O(\delta \lambda^3) \,,
\label{7-3}
\eea
where $H_0$ is the initial Hamiltonian, $|0,\lzero\rangle$ is the ground state of the initial Hamiltonian, and $\ce_0$ denotes the initial ground state energy density. Here we have used space translation invariance as well as the fact that one point functions in the initial ground state are constants in both space and time. On the other hand, the expectation value of the operator $\calo$ is, to order $\delta \lambda$, given by
\ben
\langle 0,\lambda_0|\calo (\vx,t) |0,\lambda_0 \rangle = \langle 0,\lzero| \calo (\vec{0},0) |0,\lzero \rangle -\,i \int d^{d-1}x \int_0^t dt^\prime \, \lambda(t^\prime) \, \langle 0,\lzero|[ \calo(\vx,t), \calo (\vec{0},t^\prime)] |0,\lzero \rangle \,.
\label{7-4}
\een
Using this, it is straightforward to verify that the Ward identity
\ben
\frac{d\ce (t)}{dt} = - \frac{d\lambda(t)}{dt} \, \langle 0,\lambda_0|\calo (\vx,t) |0,\lambda_0 \rangle \,,
\een
is satisfied.\footnote{ The time derivatives of the second and third lines in eq. (\ref{7-3}) cancel the time derivatives which act on the upper limit of integration in the third term of the first line.}

We are interested in evaluating eq. (\ref{7-3}) at late times. However since the coupling is a constant for $t \geq \dt$, the energy density at infinitely late times is exactly the same as the energy density at $t = \dt$. The ground state energy density of the final Hamiltonian is given by the standard expression
\ben
\ce_f = \ce_0 - \delta \lambda \, \langle 0,\lzero| \calo (\vec{0},0) |0,\lzero \rangle - \frac{\delta \lambda^2}{V^2} \sum_{n\neq 0} \frac{ |\langle 0,\lzero|\int d^{d-1}x \, \calo (\vec{x},0) |n,\lzero \rangle|^2}{\ce_0 - \ce_n} + \cdots \,,
\label{7-5}
\een
where $\ce_n$ denote the energy densities of the excited states $|n,\lambda_0\rangle$ of the initial Hamiltonian and $V$ is the volume of the system. It is clear from eqs. (\ref{7-3}) and (\ref{7-5}) that the excess energy density $\Delta \ce = \ce (\dt) - \ce_f$ starts at $O(\delta \lambda^2)$. Moreover we expect that the UV divergences in $\ce(t)$ for $t \geq \dt$ are cancelled by those in $\ce_f$. This expectation comes from the following fact: as seen in the previous sections, and in \cite{dgm1,dgm2}, the UV divergent terms depend on both $\lambda(t)$ and its time derivatives. However for $t > \dt$, the coupling is constant and these time derivatives vanish. Therefore, the UV divergent terms should be those of the constant coupling interacting theory. While this is explicit in the free field theory considered in section \ref{energylate}, we do not have an explicit proof for general interacting theories but it stands as a reasonable expectation. If the resulting expression for $\Delta \ce$ is also IR finite, it is determined to this order entirely by dimensional analysis,
\ben
\Delta \ce \sim \delta \lambda \, \dt^{d-2\Delta} \,.
\een
Further, as discussed in \cite{dgm1,dgm2}, the corrections to this result would be a power series in the dimensionless coupling $g = \delta \lambda \, \dt^{d-\Delta}$, which is small in the fast quench limit.

This argument will fail if the integrals involved in $\Delta \ce$ are IR divergent. This can be seen to happen when $2\Delta > d$, as we have explicitly seen for the free field theory for $d=2$ and $3$.

\section{Discussion}
\label{discuss}
The aim of this paper is to establish a precise relation between the
smooth fast quench and the instantaneous quench. These are the most common quench
protocols discussed in high energy theory and condensed matter physics literature,
respectively. Naive reasoning would say that if one considers the
evolution at very late times (with respect to the quench rate $\dt$),
then both protocols should give the same result since $\dt$ would be
negligible. However, our results in this paper suggest that they may
or may {\it{not}} give the same answers depending on a variety of factors such
as the spacetime dimension, the scaling dimension of the quenched operator and how much time
after the quench is considered. In this paper, we computed spatial
correlators, local expectation values and the energy density. In this
section, we summarize our results, making precise statements on when
the abrupt approximation makes sense. We also discuss different
procedures to regulate the energy density. The bulk of our comments
below relate to the explicit calculations performed in the free field
theory. However, as we will discuss at the end of this section, we
expect that these conclusions hold for generic interacting theories
for smooth fast quenches as defined in eq. (\ref{0-2}).

\subsection*{Spatial Correlators}
The study of spatial two-point correlation functions is interesting because they are UV-finite quantities that introduce a new scale to the problem, \ie the spatial separation $r$. The behaviour of this object is very different depending on the time at which we compute it. We summarize  the results here for early times, \ie $t/\dt \sim O(1)$, and for very late times, \ie $t/\dt \gg 1$ and $m t \gg1$. 

\begin{itemize}
\item At early times, we can distinguish between three different regimes. 

When $m \dt >1$, independently of $r$, we are in the slow quench regime, so we cannot compare with our previous results. However, there should be some signatures of universal behaviour corresponding to the Kibble-Zurek scaling when quenching through a critical point. We leave this appealing point for further research in the future \cite{dgm3}.

The most interesting feature appears when $r<\dt<1/m$. This means that the quench is fast since $m \dt < 1$ but the spatial separation defines the smallest scale. Then we found that the {\it{same}} universal scaling that was reported in \cite{dgm1, dgm2} appears in the two-point spatial correlation function. This holds in any spacetime dimension and for arbitrary ``early" times. We can think of this correlator as a version of $\vev{\phi^2}$ regulated with point splitting and so, $r$ plays the role of a short distance cut-off --- see below.

In general, we would continue to decrease $\dt$ towards the instantaneous limit in which $\dt \to0$. However, in this correlator, we are limited by the distance separation $r$. In fact, the correlator saturates as $\dt$ gets of order $r$ and then the result becomes independent of $\dt$. In all this analysis we were able to take the UV cut-off to infinity but we expect a similar behaviour when working in theories with a finite cut-off, with $r^{-1}$ playing that role here.

\item At late times, the results depend on the spacetime dimensions and on the separation distance. For long distances, in any dimensions, the correlator for the instantaneous quench and the smooth quench coincide. As the separation becomes smaller, the behaviour is different depending on the spacetime dimensions: for $d=3$, the smooth and the instantaneous correlator continue to coincide as $r\to0$; for $d=5$, there appears a small finite difference that goes to zero as $\dt$ goes to zero; finally, for $d=7$ the two correlators differ by an infinite amount as $r\to0$. We expect the latter behaviour extends in higher dimensions.

\end{itemize}

\subsection*{Expectation value of $\phi^2$}

We showed that the short distance expansion of the correlator is in one-to-one correspondence with the counterterms needed to regulate the bare expectation value of $\phi^2$. Then, it should not be a surprise that the behaviour of the $\vev{\phi^2}$ at late times is very similar to that of the spatial two-point correlator at late times but with small spatial separation. In fact, we showed that for $d=3$, the smooth and the instantaneous quench give the same answer. When evaluated for $d=5$, the smooth quench differs from the instantaneous quench but only in a finite amount that is of order $\dt$. In higher dimensions, however, it is impossible to regulate the expectation value of $\phi^2$ in the case of the instantaneous quench. The smooth quench, in contrast, has a smooth finite limit as $\dt \to0$ and so, the two approaches yield infinitely different results.

\subsection*{Regulating the energy density}

Both in the present and previous \cite{dgm1, dgm2} works, we worked in
a framework where UV-divergent quantities where regulated by adding
suitable counterterms. Of course, we showed how to construct those
counterterms and how they yield finite values for quantities such as
the expectation value of $\vev{\phi^2}$ and the energy density. The
way in which such a subtraction is done is much in the spirit of how
regularization works in AdS/CFT through what is known as
holographic renormalization \cite{holoren1,holoren2,holoren3}. There
we add an extra counterterm (boundary) action to the usual
gravitational action to get finite expectation values. This is how,
for instance, expectation values for holographic quantum quenches are regulated in \cite{numer}, which
served as a motivation to our studies. Our approach is also reminiscent of the way field
theories in curved spacetimes are regulated.

However, a number of things about this procedure may appear strange to a typical field theorist. In particular, the fact that our counterterms first, depend on time and second, some terms depend on time derivatives of the quenched coupling. In this section, we would like to go back to this procedure and compare it with other candidates.

To summarize, we define a renormalized energy density by subtracting counterterm contributions from a bare energy density (and taking the cut-off to infinity). Basically, $ \vev{{\cal{E}}}_{ren} \equiv {\cal{E}}_{quench} - {\cal{E}}_{ct}$ where ${\cal{E}}_{quench}$ and ${\cal{E}}_{ct}$ separately diverge as the cut-off goes to infinity but $\vev{{\cal{E}}}_{ren}$ is finite. We also showed in \cite{dgm2} that with this definition of renormalized energy density (and an equivalent for the scalar field), we satisfy the Ward identity in any spacetime dimension and for any quench protocol.

A second, perhaps more standard approach, would be to recognize the divergences as coming from the zero-point energy for the scalar field. Each momentum mode behaves as a single harmonic oscillator and then if we sum all the zero-point energies, \ie $\frac{1}{2} \omega(k)$, we get a divergent quantity. This is what we call ${\cal{E}}_{fixed}$, that is, the energy density for a scalar field of fixed mass at any instant of time.  Again, one would naively say that we should get a UV finite value if we subtract ${\cal{E}}_{quench}-{\cal{E}}_{fixed}$. One nice thing about this is that if we go to very early or very late times where the mass is constant, there is a ground state and it has precisely zero energy density and any other state has a positive energy density. Even if this procedure works for low dimensional spacetimes, we showed that is not enough in higher dimension. In fact, for $d=3$ and $5$, we have
\bea
{\cal{E}}_{fixed} - {\cal{E}}_{ct} \propto m^{d},\ \ \ \ \ \ 
\eea
while in $d=7$,
\beq
{\cal{E}}_{fixed} - {\cal{E}}_{ct} \propto m^{7} + m^{2} \, \partial_t^2 m^2 \, \Lambda,
\eeq
where $\Lambda$ is some energy UV cut-off. So first thing to note is that for $d<7$ (actually, for $d<6$) both the counterterm energy density and the fixed energy density only differ by a finite amount, so if one is sufficient to regulate the theory then, so is the other. Moreover, as most of our study corresponds to the fast quench regime where $m \dt \ll 1$, this finite amount would be negligible compared to the scaling with $\dt$ and so, the conclusion will be unchanged using either approach.\footnote{See, however, the discussion on the reverse quench in $d=3$.} Let us also note that the subtraction of ${\cal{E}}_{fixed}$ also satisfies the Ward identity. This is easy to see as the only time dependence is on the mass, so 
\beq
\partial_t {\cal{E}}_{fixed} = \partial_t \left( \sigma_s^{-1} \int dk \, k^{d-2} \sqrt{k^2+m^2(t)} \right) = \frac{1}{2} \left(\sigma_s^{-1} \int dk \,  \frac{k^{d-2}}{\sqrt{k^2+m^2(t)}} \right) \partial_t m^2(t),
\eeq
but the term in parentheses in the final expression is just the bare expectation value of $\phi^2$ with an instantaneous mass $m(t)$ --- compare to eq. (\ref{phi_sq_fixed})---, which gives exactly the Ward identity in eq. (\ref{ward_identity}).

The situation is completely different in higher dimensions, though. The energy density of the quench has more divergences than those appearing in ${\cal{E}}_{fixed}$. These are proportional to time derivatives of the quenched coupling. If we suppose that all the quench happens within a time of order $\dt$ of, say, $t=0$, then these terms will not affect what happens at very early and very late times, so it is possible to compute ${\cal{E}}_{quench}-{\cal{E}}_{fixed}$ in those regimes. However, if we want to follow the evolution through the actual quench, ${\cal{E}}_{quench}-{\cal{E}}_{fixed}$ is just divergent and we do not have a finite observable in the middle of the process. This is the main reason why ${\cal{E}}_{ren}$ is a better measure of what is going on during the quench, \ie because it allows us to compute the energy density at any time in any spacetime dimensions.

\begin{figure}[H]
   % \vspace*{-0.1cm}
    \makebox[\linewidth]{
        \includegraphics[scale=0.75]{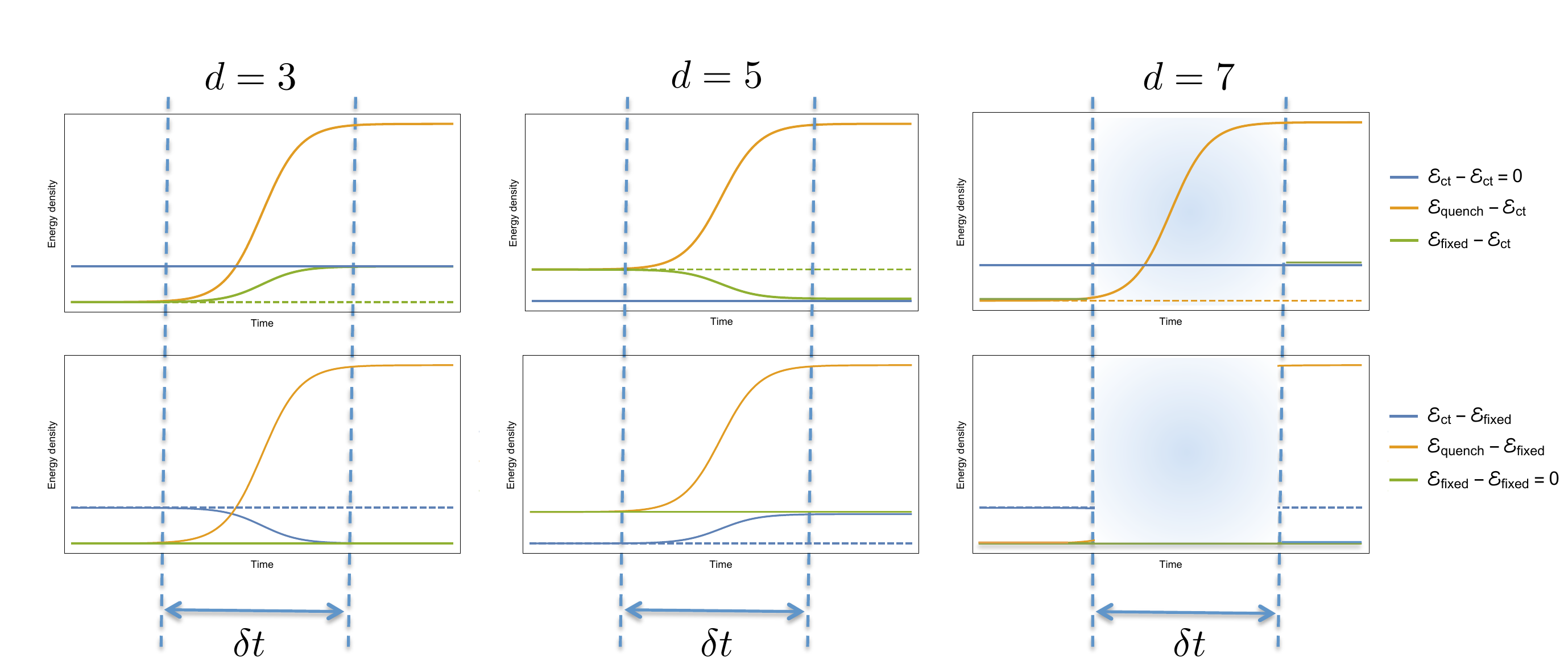}
    }
    \caption{(Colour online) Schematic description of the evolution of
      the energy density after a quench as a function of time
      depending on the regularization scheme. In the first row we
      exemplify the counterterm subtraction that allows us to follow
      the evolution for any time, even during the quench itself,
      characterized by the time scale $\dt$. Taking the counterterm
      energy density as the zero of energy density, at very early times, the starting
      energy density is negative for $d=4k+3$ and positive for $d=4k+1$, with
      $k \geq 0$ integer. In the second row we take the fixed energy density
      as the zero of energy density. In this case, all the quench energies are
      greater than this ground zero energy density. The reason is that the
      fixed mass energy density is the energy density for the scalar with a fixed mass
      and no quench, so any energy inserted during the quench will
      give a positive additional contribution. Note that in low
      dimensional spacetimes, the fixed energy density differs from the
      counterterm energy density but always by a finite amount proportional to
      a power of the mass at that instant of time, while the quench energy density usually scales also with $\dt$, so our universal scalings, \ie $\delta{\cal E} \sim \delta \lambda^2 \, \delta t^{d-2\Delta}$, will appear in any case. The same will happen in greater dimensions at very early and very late times. However, for $d\geq 6$ during the quench, it is not sufficient to subtract the fixed energy density, as there are extra UV divergences in the quenched energy density that are proportional to time derivatives of the mass. In this case, subtracting the fixed energy density is sufficient to compute the energy density at very early or very late times but not during the middle of the quench. This is depicted in the plot in the bottom-right corner for $d=7$ but is a general feature of higher dimensional spacetimes. In the same way, in the upper-right plot we cannot sketch ${\cal E}_{fixed}-{\cal E}_{ct}$ during the quench as they differ by an infinite amount.}
    \label{no_image}
\end{figure}

All of these situations are depicted in fig.~\ref{no_image}, where to
show clearly what is going on we set as the zero of energy density with ${\cal{E}}_{ct}$ in the first row and with ${\cal{E}}_{fixed}$ in the second. For $d=3$ and $5$, both approaches are valid and we see that the only difference is on some small quantity proportional to $m$. We showed both $d=3$ and $d=5$ because ${\cal{E}}_{fixed}$ and ${\cal{E}}_{ct}$ differ in each case by a different amount. While in $d=3$, the difference is negative, in $d=5$, it is positive. In $d=7$, however, we can follow ${\cal{E}}_{ren}$ but not the other one. Then, it is clear now that the naive intuition is wrong or at least is not complete.

The next step would be to think whether there is some other way to regulate the energy density in higher dimensions. A possible answer already appeared in \cite{dgm1, dgm2} while not completely emphasized. We know that the counterterms come from an adiabatic expansion. At zeroth order, this adiabatic expansion gives just the fixed energy density that corresponds to doing the quench infinitely slowly so that, at each instant of time, the energy density is just the energy density needed for the scalar field to have that particular mass $m(t)$. What we do to get the counterterms in just to expand the adiabatic expansion for large momentum and then extract the divergent pieces. As mentioned, at zeroth order, the fixed energy density differs from the counterterm energy density by a finite piece proportional to $m^d(t)$. But we also know that for higher dimensional spacetimes we need to go to higher orders in the adiabatic expansion to capture the divergences involving time derivatives of the mass. So an idea to generalize the fixed energy density subtraction to higher dimensions would be that, instead of subtracting just the counterterms, to subtract the full energy density in the adiabatic expansion to that order. This would correspond to the energy of a slow quench but going beyond the zeroth order. 

To be more explicit, in the adiabatic expansion shown in \cite{dgm2} we defined incoming modes of the form
\beq
u_\vk = \frac{1}{\sqrt{2\, \Omega_k(t)}} \exp \left(i \, \vk\cdot\vec{x} -i \int^t \Omega_k(t') dt' \right)\,.
\label{outbk}
\eeq
Then we found that
\beq
\Omega_k = \omega_k -\frac{1}{4 \omega_k} \left(\frac{\ddot{\omega}_k}{\omega_k} -\frac{3 \dot{\omega}_k^2}{2\omega_k^2} \right) + O \left(\frac{1}{m^3 \dt^4}\right),
\eeq
where $\omega_k^2=k^2+m(t)^2$. So the zeroth order term in $\Omega_k$ is the fixed mode energy density but this is then corrected with a second term that is second order in time derivatives of the mass and so forth. 

For $d=7$, we showed that it is enough to expand the energy density to that order. So instead of regulating the energy density with the counterterm energy density what we can do is to subtract, at any $t/\dt$, the whole energy density coming from that second order expansion. This will include, of course, the necessary terms to cancel all the UV divergences but it will probably introduce some extra finite terms, in analogy to the extra finite piece that the fixed energy density has with respect to the counterterm energy density in lower dimensions. 
Let us add that this subtraction is also consistent with the Ward identity.\footnote{To see this one should use the $u_\vk$ modes of eq. (\ref{outbk}) to compute the energy density for the scalar field,
\beq
\langle {\cal{E}} \rangle = \frac{\Omega_{d-2}}{2 (2\pi)^d} \int \frac{k^{d-2}dk}{2 \omega_{in}}\,\Big( |\partial_t u_\vk|^2+|\partial_i u_\vk|^2 + m(t)^2 |u_\vk|^2
\Big)\,.
\label{en_density}
\eeq
Upon taking the time derivative of that expression, one should get two different terms. First, a term proportional to 
\beq
\Omega_k^2 - \omega_k^2 - \frac{1}{2}  \frac{\partial_t^2{\Omega}_k}{\ \Omega_k} + \frac{3}{4} \left(\frac{\partial_t{\Omega}_k}{\Omega_k} \right)^2 \,, \label{Omega}
\eeq
that perfectly vanishes since that is the equation which $\Omega_k$ should satisfy in order for the modes to satisfy the equations of motion (see \cite{dgm2}). The second term, however, is not vanishing and gives exactly the Ward identity.}

In all, we have two different consistent ways of regulating the energy density in our scalar field quenches. It would be interesting to consider whether there is some analogy of these two methods in holography. Our counterterm subtraction is clearly the counterpart of the holographic renormalization approach. But what would be the equivalent of subtracting the fixed mass energy density in an holographic setup? Well, it seems reminiscent of the old method of background subtraction in the early days of the AdS/CFT correspondence (see, for instance, \cite{Gubser:1998nz}). Usually, divergences in holography appear as we take limits toward the boundary due to the divergent nature of pure anti-de Sitter spacetime as we take the radial coordinate towards the boundary. So, the first idea in holography, which was inherited from early semi-classical calculations in quantum gravity, to get a finite renormalized quantity for some excited state was to subtract that same quantity but in the vacuum state, \ie in pure AdS. Then both quantities will be divergent but their subtraction would be finite and this is quite analogous to our fixed energy density subtraction. Later on, this procedure was replaced by the more rigorous method of holographic renormalization.

\

\begin{figure}[H]
        \centering
        \subfigure[The energy density using the counterterm subtraction]{
                \includegraphics[scale=1]{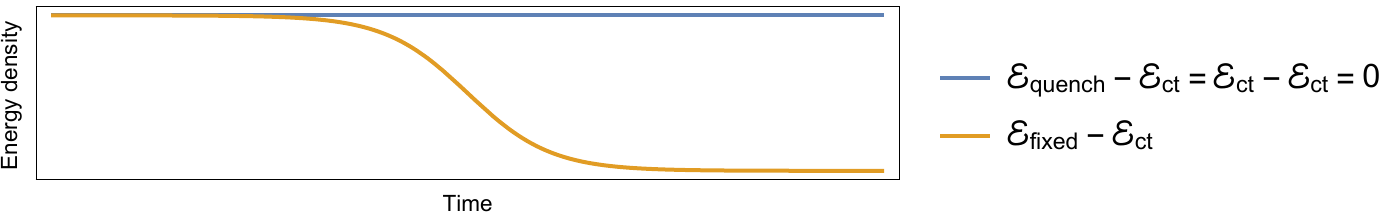} \label{fig_d3_ener_1}}
        \subfigure[The energy density using the fixed mass subtraction]{
                \includegraphics[scale=1]{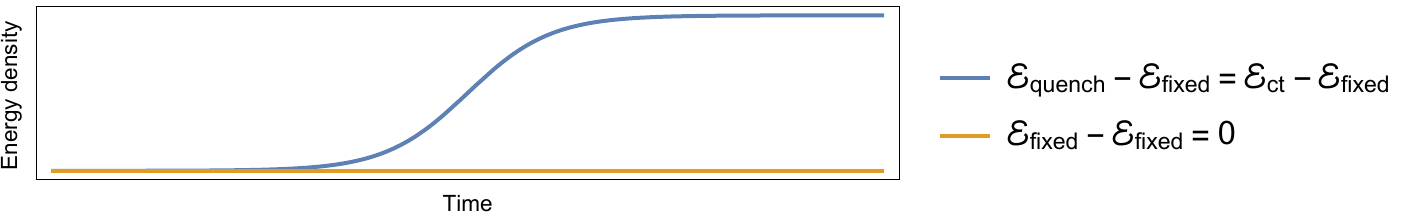} \label{fig_d3_ener_2}}
        \caption{(Colour online) Schematic plots of the evolution of the energy density in three spacetime dimensions when $\dt \to0$. From the counterterm subtraction point of view, the work done is zero but from the fixed mass subtraction perspective, there is finite work done.}
                \label{fig_d3ener}
\end{figure}

There is an interesting effect in the use of these two different approaches in $d=3$. In this case, the scaling is special because instead of giving a diverging behaviour as $\dt \to0$, it gives a vanishing one. We analysed this case in \cite{dgm2}, concluding that actually the energy density produced was given by $\delta\langle {\cal{E}} \rangle_{ren} =\frac{m^3}{8\pi}$, where $m$ is the initial mass. The interesting thing was that then, if we do the reverse quench where the initial mass is zero, it appears that the work done by the quench is zero! However, this was an artifact of using the counterterm energy density to regulate the expectation value. If we use the fixed energy density then we will find that the energy density starts from zero at early times and it goes to some finite value, giving some non-zero finite work in the process, as depicted in fig.~\ref{fig_d3ener}.

\

Finally, we wish to consider one last method of obtaining a finite energy density. In section \ref{energylate}, we considered the difference between two physical energies, \ie ${\cal{E}}_{quench} - {\cal{E}}_{ground}$. Note that ${\cal{E}}_{ground}$, the ground-state energy density, should be something that we can easily define at very late and very early times. In particular, this looks like ${\cal{E}}_{quench} - {\cal{E}}_{fixed}$ at these early and late times. However, ${\cal{E}}_{ground}$ is a real physical energy density of a particular state and it can be defined in any renormalization scheme. So in this case, we do not need to make any reference to our choice of scheme because all of the divergences cancel in the difference of two physical energies. Of course, the drawback is that this can only be computed at very early or late times.

\subsection*{Even dimensions}

Most of the explicit calculations presented in this paper refer to odd spacetime dimensions. However, most of the conclusions also hold for even dimensions. As pointed out in \cite{dgm1,dgm2}, the only differences between even and odd dimensions are that there are additional logarithmic UV divergences which must be regulated in even dimensions, and as a result, the renormalized expectation values have an extra logarithmic enhancement in the $\dt$ scaling. For example, in even dimensions, the expectation value for $\phi^2$ in the fast smooth quench scales as
\begin{equation}
\vev{\phi^2}_{ren} \sim \dt^{4-d} \log \mu \dt,
\end{equation} 
where $\mu$ is a new renormalization scale introduced by the logarithmic counterterms. 

The appearance of logarithmic counterterms adds an additional technical difficulty to the calculations presented in this paper but we do not expect that they would change the main results. In particular, the renormalized expectation values of $\phi^2$ have a smooth limit at late times as $\dt \to0$ in higher even or odd dimensions, while the analogous quantity diverges after an instantaneous quench. With regards to the energy density, we presented results for lower even dimensions in section \ref{energylate}, where we showed that the excess energy has a smooth limit in $d=2$ as $\dt \to0$, which matches the instantaneous answer. In higher dimensions, the excess energy diverges as expected from the scaling of $\vev{\phi^2}_{ren}$ and the Ward identity \reef{ward_identity}.

\subsection*{Lessons for interacting theories}

We end this discussion with a comment on the lessons of our work for
general interacting theories. 
In \cite{dgm1,dgm2} we showed that the
scaling form of renormalized quantities holds for general quantum
field theories for fast quenches as defined in eq. (\ref{0-1}). It is
natural to expect that the scaling for correlation functions found for
the free theory in section 4 would have an analogue in interacting
theories as well. When the length scale in the correlator is small
compared to the quench time, this correlator can be viewed as a point-split version of the operator which is used for the quench and in the
fast quench limit, the arguments of \cite{dgm1,dgm2} then show that this
quantity would scale in the expected fashion.

In this paper, we found that the relationship between the fast limit of
a smooth quench and an instantaneous quench is non trivial for free
field theories in high dimensions. This again should generalize to
interacting theories. What really led to the non trivial relation is
the fact that in higher spacetime dimensions, the conformal dimension
of the quenched operator becomes large. Indeed, the scaling of the
renormalized quenched operator $\calo$ for general interacting theory
together with the Ward identity shows that the renormalized energy density at
late times behaves as $\dt^{d-2\Delta}$ and therefore diverges as $\dt\to0$ for any
$d$ whenever $\Delta > d/2$. This can be consistent with the results
of an instantaneous quench only if the latter is UV divergent in this
case. This fact should have non trivial consequences for the ability
to express the state after a quench in terms of a boundary state as in
\cite{cc2} even in low spacetime dimensions when the conformal
dimension of the quenched operator is large enough.

\section*{Acknowledgements} 
We would like to thank Diptarka Das, Fabian Essler,  Alfred Shapere, Stephen Shenker, Eva Silverstein and Lenny Susskind for discussions.
S.R.D. would like to thank the Galileo Galilei Institute for Theoretical Physics for the hospitality and the INFN for partial support during the completion of this work.
Research at Perimeter Institute is supported by the
Government of Canada through Industry Canada and by the Province of Ontario
through the Ministry of Research \& Innovation. RCM and DAG are also supported
by an NSERC Discovery grant. RCM is also supported by research funding
from the Canadian Institute for Advanced Research. 
The work of SRD is partially supported by the National Science Foundation grant NSF-PHY-1214341. DAG also thanks the Kavli Institute for Theoretical Physics for hospitality during the last stages of this project. Research at KITP is supported, in part, by the National Science Foundation under Grant No. NSF PHY11-25915.

\appendix
\section{Review of constant mass correlators}
\label{const_mass_corr}

In this appendix, we review the computation and behaviour of the (spatial) correlator for a massive free scalar field with a constant mass. For simplicity, we will focus on odd spacetime dimensions. Hence we are interested in computing the following spatial correlator,
\beq
C(\vec{r}) \equiv \langle \phi(\vec{r}) \phi(\vec{0}) \rangle = \frac{1}{2 (2\pi)^{d-1}}\int \frac{d^{d-1}k}{\sqrt{k^2+m^2}}\, e^{i \vec{k} \cdot \vec{r}},
\label{sp_corr}
\eeq
where $m$ is simply a fixed constant (for all time). First, we can choose, without loss of generality, to place $\vec{r}$ along one particular axis using the rotational symmetry
of the problem. Integrating out the transverse angular directions, then yields
\beq
C(\vec{r}) = \frac{\Omega_{d-3}}{2 (2\pi)^{d-1}} \int \frac{k^{d-2}dk}{\sqrt{k^2+m^2}} \int_0^\pi d\theta \sin^{d-3} \theta e^{i k r \cos \theta},
\eeq
where $k=|\vec{k}|$ and $r=|\vec{r}|$. The integral over $\theta$ can be done analytically and for odd $d$, we find
\beq
C(\vec{r}) =  \frac{1}{\sigma_c\,r^{\frac{d-3}{2}}} \int \frac{k^{\frac{d-1}{2}}dk}{\sqrt{k^2+m^2}}\, J_{\frac{d-3}{2}}(k r)\,,
\label{mass_corr}
\eeq
where $\sigma_c=2^{\frac{d+1}{2}}\pi^{\frac{d-1}{2}}$ and $J_{\frac{d-3}{2}}$ is the Bessel function of order $\frac{d-3}{2}$. 

Now, to get the full answer for $C(\vec{r})$ we need to integrate over all $k$, so we note that for large $k$, the Bessel function behaves as $1/\sqrt{kr}$ times some linear combination of trigonometric functions (of $kr$). Hence the na\"ive counting of the powers of $k$ would yield an overall factor of $k^{\frac{d-4}2}$ in the integrand above and hence one might conclude that the integral would diverge for any $d\ (\ge2)$. However, this factor provides a envelope for a rapidly oscillating function which tends to produce an added cancellation in the integral. Integrals of this form can be defined with the following regulator: Insert an additional factor of the form $\exp(- a k)$ to the desired integrand. The resulting (finite) answer now remains finite in the limit of $a\to0$.  In fact, one can show that this method works correctly for integrands that are a product of some power of $k$ times an oscillatory function \textit{around zero}. For example, one can show that,
\begin{eqnarray}
\int_0^\infty dk\, k^\alpha \sin (x k+\delta) & \equiv & \lim_{a \to0} \int_0^\infty dk\, k^\alpha \sin (x k+\delta) \exp(- a k) \nnn\\
 &=& \sin \left(\frac{\pi}{2} (\alpha +1)+\delta\right) \Gamma (\alpha +1)\, x^{-(\alpha +1)}\,, \label{trig_reg} 
 %\\
%\int_0^\infty dk\, k^\alpha \cos (x k) & = & \lim_{a \to0} \int_0^\infty dk\, k^\alpha \cos (x k) \exp(- a k) \nnn \\
%&=& \cos \left(\frac{\pi}{2}  (\alpha +1)\right) \Gamma (\alpha +1) x^{-(\alpha +1)} \,, \nnn 
\end{eqnarray}
for any non-negative values of $\alpha$ and $x$. Applying this apprach to eq.~\reef{mass_corr} yields the following analytic answer
\beq
C(\vec{r}) = \frac{1}{(2\pi)^{d/2}}  \left(\frac{m}{r}\right)^{\frac{d-2}{2}}  K_{1-\frac{d}{2}}(m r)\,,\labell{king}
\eeq
where now $K_\alpha$ is the Bessel K function. Of course, one can readily verify that this correlator satisfies the Klein-Gordon equation as desired.\footnote{The correlator (\ref{king}) applies for general spacelike separations if we replace the distance $r$ by $\sqrt{r^2-t^2}$.}

Given this expression \reef{king}, it is straightforward to examine various asymptotics of  the correlator.
In particular, considering the limit $mr \to0$, we obtain
\bea
\sigma_c C(\vec{r}) & = & \frac{\Gamma \left(\frac{d}{2}-1\right) 2^{\frac{d-3}{2}}}{\sqrt{\pi } r^{d-2}} \left(1 -\frac{m^2 r^2}{2(d-4)}  + \frac{m^4 r^4}{8 (d-4) (d-6)} +O\left(m^6 r^6\right)\right) + \nnn \\
& & +\frac{ \Gamma \left(1-\frac{d}{2}\right)}{\sqrt{\pi } 2^{\frac{d-1}{2}}} m^{d-2} + O(m^d r^2)\,.
\label{div_corr}
\eea
Hence the correlator diverges as expected as $r\to0$, \ie the leading divergence goes as $1/r^{d-2}$. The above expansion reveals that the subleading terms are all proportional to the mass in this limit. To have some concrete examples, we show: 
\begin{eqnarray}
d & = & 3, \hspace{1cm} \sigma_c \, C(\vec{r}) = \frac{1}{r}-m + O(m^2 r)\,, \nnn\\
d & = & 5, \hspace{1cm} \sigma_c \, C(\vec{r}) = \frac{1}{r^3} -\frac{m^2}{2 r} + \frac{m^3}{3} + O(m^4 r)\,, \label{d7_div_corr}\\
d & = & 7, \hspace{1cm} \sigma_c \, C(\vec{r}) = \frac{3}{r^5} -\frac{m^2}{2 r^3} +\frac{m^4}{8 r} -\frac{m^5}{15}+ O(m^6 r)\,.\nnn
\end{eqnarray}
 We also note that the leading term in eq.~\reef{div_corr} is, in fact, the exact answer for the massless correlator, \ie
\beq
C(\vec{r}) = \frac{\Gamma \left({ \frac{d-2}{2}}\right) }{4\pi^{\frac{d}{2}}}\ \frac{1}{ r^{d-2}}  
\qquad{\rm for}\ m=0\,.
\label{nomass}
\eeq

The other interesting limit to consider is $m r \to\infty$. In this case, the Bessel function decays exponentially and we find 
\beq
C(\vec{r}) = \frac{e^{-m r}}{\sigma_c} \, \frac{m^{\frac{d-3}{2}}}{r^{\frac{d-1}{2}}}\, \left( 1+ O(1/(mr)) \right)\,.
\label{largerad}
\eeq
We might note that the power of $r$ in the leading term in eqs.~\reef{div_corr} and \reef{largerad} happens to coincide for $d=3$ but otherwise they differ.

To conclude this appendix, we emphasize the two main results: The first one is that na\"ively the integrals above seem to be divergent, especially for high $d$. However, because the integrand is mainly oscillating around zero, they can be regulated as in eq.~\reef{trig_reg} to get a finite result. The second lesson is that in the static case this correlator diverges as $1/r^{d-2}$, as shown in eqs.~\reef{div_corr} and \reef{d7_div_corr}. We will take these facts into account when we analyse spatial correlators in the instantaneous and smooth quenches in section \ref{late}.

\end{document}